\documentclass{supp}
\usepackage{graphicx,pdfpages}

\title{Supplementary Information}

\spacing{1}

\begin{document}
 

# The Transiting Circumbinary Planets Kepler-34 and Kepler-35

William F. Welsh[1], Jerome A. Orosz[1], Joshua A. Carter[2], Daniel C. Fabrycky[3], Eric B. Ford[4], Jack J. Lissauer[5], Andrej Prsa[6], Samuel N. Quinn[2,22], Darin Ragozzine[2], Donald R. Short[1], Guillermo Torres[2], Joshua N. Winn[7], Laurance R. Doyle[8], Thomas Barclay[5,19], Natalie Batalha[5,20], Steven Bloemen[23], Erik Brugamyer[9], Lars A. Buchhave[10,21], Caroline Caldwell[9], Douglas A. Caldwell[8], Jessie L. Christiansen[5,8], David R. Ciardi[11], William D. Cochran[9], Michael Endl[9], Jonathan J. Fortney[12], Thomas N. Gautier III[13], Ronald L. Gilliland[25], Michael R. Haas[5], Jennifer R. Hall[24], Matthew J. Holman[2], Andrew W. Howard[14], Steve B. Howell[5], Howard Isaacson[14], Jon M. Jenkins[5,8], Todd C. Klaus[24], David W. Latham[2], Jie Li[5,8], Geoffrey W. Marcy[14], Tsevi Mazeh[15], Elisa V. Quintana[5,8], Paul Robertson[9], Avi Shporer[16,18], Jason H. Steffen[17], Gur Windmiller[1], David G. Koch[5], and William J. Borucki[5]

[1]Astronomy Department, San Diego State University, 5500 Campanile Dr. San Diego, CA 92182, USA
[2]Harvard-Smithsonian Center for Astrophysics, 60 Garden Street, Cambridge, MA 02138, USA
[3]UCO/Lick Observatory, University of California, Santa Cruz, CA 95064, USA
[4]University of Florida, 211 Bryant Space Science Center, Gainesville, FL 32611-2055, USA
[5]NASA Ames Research Center, Moffett Field, CA, 94035, USA
[6]Villanova Univ., Dept. of Astronomy and Astrophysics, 800 E Lancaster Ave, Villanova, PA 19085, USA
[7]Massachusetts Institute of Technology, Physics Department and Kavli Institute for Astrophysics and Space Research, 77 Massachusetts Avenue, Cambridge, MA 02139, USA
[8]Carl Sagan Center for the Study of Life in the Universe, SETI Institute, 189 Bernardo Avenue, Mountain View, CA 94043, USA
[9]McDonald Observatory, The University of Texas at Austin, Austin TX 78712-0259, USA
[10]Niels Bohr Institute, Copenhagen University, Juliane Maries Vej 30, DK-2100 Copenhagen, Denmark
[11] NASA Exoplanet Science Institute/Caltech, 770 South Wilson Ave, Pasadena, CA USA 91125, USA
[12]Dept. of Astronomy and Astrophysics, Univ. of California, Santa Cruz, Santa Cruz, CA 95064, USA
[13]Jet Propulsion Laboratory, 4800 Oak Grove Drive, Pasadena, CA 91109, USA
[14]Astronomy Department, University of California, Berkeley, CA, 94720, USA
[15]School of Physics and Astronomy, Tel Aviv University, Tel Aviv 69978, Israel
[16]Las Cumbres Observatory Global Telescope Network, 6740 Cortona Dr., Ste 102, Santa Barbara, CA 93117, USA
[17]Fermilab Center for Particle Astrophysics, MS 127, PO Box 500, Batavia, IL 60510, USA
[18] Department of Physics, Broida Hall, University of California, Santa Barbara, CA 93106, USA
[19] Bay Area Environmental Research Institute, Inc., 560 Third St. West, Sonoma, CA 95476, USA
[20]Dept of Physics & Astronomy, San Jose State Univ., One Washington Square, San Jose, CA 95192, USA
[21]Centre for Star and Planet Formation, Natural History Museum of Denmark, University of Copenhagen, DK-1350 Copenhagen, Denmark
[22]Department of Physics & Astronomy, Georgia State University, PO Box 4106, Atlanta, GA 30302, USA
[23]Instituut voor Sterrenkunde, Katholieke Universiteit Leuven, Celestijnenlaan 200D, B-3001 Leuven, Belgium
[24]Orbital Sciences Corporation/NASA Ames Research Center, Moffett Field, CA 94035
[25]Space Telescope Science Institute, Baltimore, MD 21218, USA

**Most Sun-like stars in the Galaxy reside in gravitationally-bound pairs of stars called "binary stars"[1,2]. While long anticipated[3-8], the existence of a "circumbinary planet" orbiting such a pair of normal stars was not definitively established until the discovery[9] of Kepler-16. Incontrovertible evidence was provided by the miniature eclipses ("transits") of the stars by the planet. However, questions remain about the prevalence of circumbinary planets and their range of orbital and physical properties. Here we present two additional transiting circumbinary planets, Kepler-34 and Kepler-35. Each is a low-density gas giant planet on an orbit closely aligned with that of its parent stars. Kepler-34 orbits two Sun-like stars every 289 days, while Kepler-35 orbits a pair of smaller stars (89% and 81% of the Sun's mass) every 131 days. Due to the orbital motion of the stars, the planets experience large multi-periodic variations in incident stellar radiation. The observed rate of circumbinary planets implies > ~1% of close binary stars have giant planets in nearly coplanar orbits, yielding a Galactic population of at least several million.**

The new planets were identified using 671 days of data from the NASA *Kepler* spacecraft[10]. As part of its mission[11] to detect Earth-like planets via the transit method, *Kepler* is monitoring over 2,000 eclipsing binary stars[12,13]. From these we selected a sample of 750 systems with orbital periods ranging from 0.9 to 276 days, and for which eclipses of both stars occur. For each system, we measured the eclipse times and searched for departures from strict periodicity, as would be produced by gravitational perturbations from a third body.

All 750 systems were searched by eye for planetary transits, with particular attention to an 18% subset that exhibited significant differences between the periods derived from the deeper primary eclipses, and those from the shallower secondary eclipses (for details see the Supplementary Information, SI). This led to the discovery of Kepler-34 and Kepler-35, and a candidate system KOI-2939. KOI-2939 (Kepler Input Catalog[14] number 05473556) exhibited a single transit at BJD 2,454,996.995 ± 0.010 of duration 2.5 hours and depth 0.18%. The transit duration constrains the size and velocity of the third body and is consistent with a Jovian planet transiting the secondary star, but we cannot verify its planetary nature. We defer discussion for a future investigation.

The stars of Kepler-34 have an orbital period of 28 days, with a period difference between primary and secondary eclipses of 4.91 ± 0.59 s. Three transits were detected (Fig 1), with the first and second being transits of the primary star, while the third is of the secondary star. Notably the transit durations are all different, ruling out the most common type of "false positive," a background eclipsing binary. Circumbinary transits naturally vary in duration as a consequence of the changing velocity of the stars. The *Kepler* photometry were supplemented by spectroscopic observations of the radial-velocity variations of both stars (Fig. 1f), in order to determine the orbital scale and sizes of all three bodies. The photometric and spectroscopic data

were fit with a model[9,15] that accounts for the three-body gravitational dynamics and the loss of light due to eclipses and transits (see SI). The model fit confirms that the transiting body is a planet with 22% the mass of Jupiter (69 Earth masses) and 76% the radius of Jupiter (8.6 Earth radii). The primary and secondary stars are similar to the Sun. With the spectra we also measured the effective temperature and abundance of heavy elements (metallicity) of both stars. The observed stellar parameters match the Yonsei-Yale theoretical models of stellar evolution[16] for an age of 5-6 Gyr. The parameters and uncertainties are given in Table 1, with details in the SI.

The stars of Kepler-35 have an orbital period of 21 days, with a period difference between primary and secondary eclipses of 1.89 ± 0.48 s. Four transits were detected (Fig. 2 b,c,d,e). The first, second, and fourth events are transits of the primary star, and the weaker third event is a transit of the secondary star. Transits do not occur every planetary orbit, placing a strong constraint on the mutual orbital inclination and its evolution. The transits differ in duration, and the interval between transits is not constant, again signaling a circumbinary body. Modeling the photometry and radial velocities yields the system parameters given in Table 1. The transiting body is a planet with 13% of the mass and 73% of the radius of Jupiter (41 Earth masses and 8.2 Earth radii). Comparison to stellar-evolutionary models suggests a system age of ~8-12 Gyr, although, interestingly, the models do not provide a satisfactory match to the stellar masses and radii under the assumption of a common age and metallicity (see SI).

The mean densities of the Kepler-34 and Kepler-35 planets are 0.61 and 0.41 g cm$^{-3}$, somewhat lower than the 0.96 g cm$^{-3}$ of Kepler-16's planet, but all are consistent with low-density gas-giant planets. Fig. 3 gives a visual comparison of the systems' orbits. For all three systems the planetary and stellar orbits are aligned to within 2 degrees, suggesting that each system formed from a flat disk of material. The period ratios (planetary to stellar) for Kepler-34, -35, and -16 are 10.4, 6.3, and 5.6, respectively, only 21%, 24%, and 14% larger than analytic estimates for stability against three-body interactions[17,18,19]. Long-term integration of the equations of motion confirms that these two new systems are stable for at least 10 Myr (SI). Note that the planets' locations bracket the habitable zone[20] (where liquid water would be stable on the surface of a rocky planet), with the Kepler-34 and Kepler-35 planets lying interior to the habitable zone and the Kepler-16 planet lying exterior.

A simple argument suggests that circumbinary giant planets are not extremely rare, as three such objects have been seen in our sample of 750 systems. Given the orbital geometry of Kepler-34, Kepler-35, and Kepler-16, the probability[21] that a randomly placed observer who sees stellar eclipses would also see planetary transits is approximately 12%, 14% and 21%, respectively (see SI). If this probability of roughly ~15% were constant across all 750 target systems, then the fraction of binaries with circumbinary gas giant planets at similar periods would be (3/750)x(0.15)$^{-1}$, or a few percent. However, this does not account for the period distribution of binaries in our sample, and the search is not complete; consequently a lower limit of ~1% is reasonable. With ~2.6% of all Sun-like stars in the Galaxy residing in binary star systems similar

to the three discussed here[2,22] (see SI), a conservative estimate yields millions of nearly coplanar circumbinary planets in the Galaxy like the ones reported here.

Orbital motion of the central stars causes complex time variations in stellar insolation for circumbinary planets. Fig. 4 shows the calculated insolation for Kepler-34 and Kepler-35. The variation is multi-periodic, with changes on the timescales of the stellar orbit, the planetary orbit, and the long-term precession of the orbits due to three-body effects. For Kepler-34 and Kepler-35, the average insolation is (respectively) 2.4 and 3.6 times the Earth's insolation, with maximum-to-minimum ratios of 250% and 160%. By comparison, for Venus the insolation is 1.9 times the Earth's with only a 2.7% variation. These highly variable, multi-periodic fluctuations in insolation are unique to circumbinary planets, and can lead to complex climate cycles. It will be interesting to explore the effects of these swings in insolation on the atmospheric dynamics (see SI), and ultimately on the evolution of life on habitable circumbinary planets.

**References:**
1. Binnendijk, L. *Properties of Double Stars*. (Philadelphia, University of Pennsylvania Press, 1960).

2. Raghavan, D. *et al*. A Survey of Stellar Families: Multiplicity of Solar-type Stars. *Astrophys. J.* **190** (Suppl.), 1-42 (2010).

3. Schneider, J. & Chevreton, M. The Photometric Search for Earth-Sized Extrasolar Planets by Occultation in Binary Systems. *Astron. Astrophys.* **232,** 251 (1990).

4. Quintana, E.V. & Lissauer, J.J. Terrestrial planet formation surrounding close binary stars. *Icarus* **185**, 1-20 (2006).

5. Deeg, H.-J. *et al*. Extrasolar Planet Detection by Binary Stellar Eclipse Timing: Evidence for the Third Body Around CM Draconis, *Astron. Astroph.* **480**, 563-571 (2008).

6. Haghighipour, N. *Planets in Binary Star Systems*. Astrophysics and Space Science Library, Vol. 366 (Springer, 2010).

7. Sybilski, P., Konacki, M. & Kozlowski, S. Detecting circumbinary planets using eclipse timing of binary stars - numerical simulations. *Mon. Not. R. Astron. Soc*. **405**, 657-665 (2011).

8. Schwarz, R., *et al.* Prospects of the detection of circumbinary planets with Kepler and CoRoT using the variations of eclipse timing. *Mon. Not. R. Astron. Soc.* **414,** 2763-2770 (2011).

9. Doyle, L.R. *et al*. Kepler-16: A Transiting Circumbinary Planet. *Science* **333,** 1602-1606 (2011).

10. Koch, D., *et al.* Kepler Mission Design. *Astroph. J.* **713**, L79-L86 (2010).

11. Borucki, W.J. *et al*. Kepler Planet-Detection Mission: Introduction and First Results. *Science* **327**, 977-980 (2010).

12. Prsa, A. *et al*. Kepler Eclipsing Binary Stars. I. Catalog and Principal Characterization of 1879 Eclipsing Binaries in the First Data Release, *Astron. J.* **141,** 83-98 (2010).

13. Slawson, R.W. *et al*. Kepler Eclipsing Binary Stars. II. 2165 Eclipsing Binaries in the Second Data Release. *Astron. J.* **142,** 160-173 (2011).

14. Brown, T.M. *et al.* Kepler Input Catalog: Photometric Calibration and Stellar Classification. *Astron. J.* **142,** 112-129 (2011).

15. Carter, J.A. *et al.* KOI-126: A Triply Eclipsing Hierarchical Triple with Two Low-Mass Stars. *Science* **331,** 562-565 (2011).

16. Yi, S.K. *et al*. Toward Better Age Estimates for Stellar Populations: The $Y^2$ Isochrones for Solar Mixture. *Astrophys J.* **136** (Suppl.), 417- 437 (2001).

17. Holman, M.J. & Wiegert, P.A. Long-Term Stability of Planets in Binary Systems. *Astron. J.* **117,** 621-628 (1999).

18. Eggleton, P. & Kiseleva, L. An Empirical Condition for Stability of Hierarchical Triple Systems. *Astrophys. J.* **455**, 640-645 (1995).

19. Doolin, S. & Blundell, K.M. The dynamics and stability of circumbinary orbits. *Mon. Not. R. Astron. Soc.,* (in the press) (2011); (arXiv:1108.4144).

20. Kasting, J.F., Whitmire, D.P. & Reynolds, R.T. Habitable Zones around Main Sequence Stars. *Icarus* **101**, 108-128 (1993).

21. Ragozzine D. & Holman, M. J. The Value of Systems with Multiple Transiting Planets. *Astrophys. J.* (submitted); preprint at arXiv:1006.3727 (2010).

22. Chabrier, G. Galactic Stellar and Substellar Initial Mass Function. *Publ. Astron. Soc. Pacif.* **809**, 763-795 (2003).

**Supplementary Information** is linked to the online version of the paper at www.nature.com/nature.

**Author Information**
W.F. Welsh, wfw@sciences.sdsu.edu
J. A. Orosz, orosz@sciences.sdsu.edu
J. A. Carter, jacarter@cfa.harvard.edu

**Author Contributions**
W.W. led the research effort on these transiting circumbinary planets (CBP), wrote much of the manuscript.
J.O. led the ETV investigation; measured O-Cs, inspected light curves, measured EB properties, measured radial velocities and flux ratios, generated Figs 1 and 2, and assembled the SI.
J.A.C. created and used the photometric-dynamical code to model the light curve and RVs; measured system parameters; generated Table 1 and Figure 3.
D.F. produced initial dynamical models to interpret the timing of eclipse and transit events leading to the planet interpretation; development of criteria for non-eclipsing CBP searches.
E.F. contributed to interpretation and text; checked long-term stability; insolation calculations.
J.J.L. contributed to interpretation and text; initiated study of variations in insolation upon CBPs.
A.P. measured mass, radii, and other properties of the EBs including contamination and flux ratios.
S.Q. obtained and analysed spectra, determined stellar parameters and luminosity ratios.
D.R. computed the estimated frequency of circumbinary planets.
D.S. developed the automated ETV code to measure eclipse times and O-C deviations.
G.T. contributed to the discussion of the stellar parameters and carried out the comparison with stellar evolution models.
J.W. contributed to the text, estimated age via gyrochronology, contributed to topics related to pseudosynchronicity.
L.D. contributed to the habitable zone discussion and spearheaded the initial search for CBPs

*The remaining authors listed below contributed significantly to the paper:*
T.B. examined pixel level data and contributed basis-vector corrected light curves.
N.B. directed EB target selection and identification.
S.B. contributed to the text and the Supplementary Information section
E.B. carried out an independent spectroscopic investigation to measure stellar parameters.
L.B. gathered spectroscopic observations for the RV and spectroscopic parameter determination.
C.C. contributed three nights of spectroscopic observations at the McDonald 2.7m observatory.
D.A.C. contributed to calibration of the *Kepler* photometer and pipeline necessary for data acquisition.
J.L.C. supported the science operations to collect and calibrate the *Kepler* data used here.
D.R.C. coordinated ground-based follow-up observations.
W.C. obtained the HET spectra, and processed all of the McDonald 2.7m and HET spectra.
M.E. contributed HET and McDonald 2.7m spectra.
J. F. contributed calculations and discussion regarding the characteristics of the planets' atmospheres.
N.G. coordinated the *Kepler* follow-up observation (KFOP) effort.
R.G. provided Mission support and contributed directly to the text and discussion of the results.
M.H. led the effort to gather, process, and distribute the data necessary for this investigation.
J.H. contributed to the collection, validation, and management of the *Kepler* data used here.
M.H. contributed to the discussion of the dynamical stability.
A.H. made spectroscopic observations using Keck-HIRES.
S.H. contributed reconnaissance spectroscopy.

H.I. obtained spectroscopic observations of targets.

J.J. developed observation/analysis techniques and calibration software that enables the *Kepler* photometer to operate successfully.

T.K. led the design and development of the Science Processing Pipeline Infrastructure needed to process the data used in this investigation.

D.L. contributed spectroscopy and preparation of the *Kepler* Input Catalog.

J.L. contributed to the development of the Data Validation component of the *Kepler* Science Operations Center pipeline necessary to obtain these data.

G.M. obtained Keck-HIRES spectra.

T.M. analysed the beaming effect in Kepler-35 and participated in the discussion of statistical inference and the spectroscopic light ratio.

E.Q. developed calibration/validation software necessary for the *Kepler* data in this paper.

P.R. contributed ten nights of spectroscopic observations at the McDonald 2.7 telescope.

A.S. contributed ground-based follow-up imaging of the targets with FTN.

J.S. contributed to the text, scope, and interpretation.

G.W. ran the ETV code, developed tools for analysing O-C variations, assisted with text.

D.K. designed major portions of the *Kepler* photometer that acquired these data.

W.B. led the design and development of the *Kepler* Mission that acquired these data, contributed to the text.

**The authors declare no competing financial interests.**

**TABLE 1: Circumbinary planet system parameters.**

| | Kepler-34 | Kepler-35 | Kepler-16 |
|---|---|---|---|
| **Planetary Properties** | | | |
| Mass of planet, $M_p(M_{\text{Jupiter}})$ | $0.220^{+0.011}_{-0.010}$ | $0.127^{+0.020}_{-0.021}$ | $0.333^{+0.016}_{-0.016}$ |
| Radius of planet, $R_p(R_{\text{Jupiter}})$ | $0.764^{+0.012}_{-0.014}$ | $0.728^{+0.014}_{-0.014}$ | $0.7538^{+0.0026}_{-0.0023}$ |
| Mean Density of planet, $\rho_p$ (g cm$^{-3}$) | $0.613^{+0.045}_{-0.041}$ | $0.409^{+0.069}_{-0.070}$ | $0.964^{+0.047}_{-0.046}$ |
| **Properties of the Planetary Orbit** | | | |
| Reference Epoch (BJD) | 2,454,969.20000 | 2,454,965.85000 | 2,455,212.12316 |
| Period, $P$ (days) | $288.822^{+0.063}_{-0.081}$ | $131.455^{+0.076}_{-0.105}$ | $228.776^{+0.020}_{-0.037}$ |
| Semi-major axis length, $a$ (AU) | $1.0896^{+0.0009}_{-0.0009}$ | $0.60345^{+0.00100}_{-0.00102}$ | $0.7048^{+0.0011}_{-0.0011}$ |
| Eccentricity, $e$ | $0.182^{+0.016}_{-0.020}$ | $0.042^{+0.007}_{-0.004}$ | $0.00685^{+0.00101}_{-0.00146}$ |
| Eccentricity Times Sine of Arg. of Periapse, $e\sin(\omega)$ | $0.025^{+0.007}_{-0.007}$ | $0.034^{+0.009}_{-0.011}$ | $-0.00448^{+0.00088}_{-0.00090}$ |
| Eccentricity Times Cosine of Arg. of Periapse, $e\cos(\omega)$ | $0.180^{+0.016}_{-0.021}$ | $0.017^{+0.021}_{-0.017}$ | $0.0051^{+0.0014}_{-0.0028}$ |
| Mean Longitude, $\lambda \equiv M + \omega$ (deg) | $106.5^{+2.5}_{-2.0}$ | $136.3^{+2.1}_{-2.7}$ | $106.51^{+0.32}_{-0.16}$ |
| Inclination $i$ (deg) | $90.355^{+0.026}_{-0.018}$ | $90.76^{+0.12}_{-0.09}$ | $90.0322^{+0.0022}_{-0.0023}$ |
| Relative Nodal Longitude, $\Omega$ (deg) | $-1.74^{+0.14}_{-0.16}$ | $-1.24^{+0.24}_{-0.33}$ | $0.003^{+0.013}_{-0.013}$ |
| **Properties of the Stars in the Stellar Binary** | | | |
| Mass of A, $M_A(M_\odot)$ | $1.0479^{+0.0033}_{-0.0030}$ | $0.8876^{+0.0051}_{-0.0053}$ | $0.6897^{+0.0035}_{-0.0034}$ |
| Radius of A, $R_A(R_\odot)$ | $1.1618^{+0.0027}_{-0.0031}$ | $1.0284^{+0.0020}_{-0.0019}$ | $0.6489^{+0.0013}_{-0.0013}$ |
| Mass of B, $M_B(M_\odot)$ | $1.0208^{+0.0022}_{-0.0022}$ | $0.8094^{+0.0041}_{-0.0044}$ | $0.20255^{+0.00066}_{-0.00065}$ |
| Radius of B, $R_B(R_\odot)$ | $1.0927^{+0.0032}_{-0.0027}$ | $0.7861^{+0.0020}_{-0.0022}$ | $0.22623^{+0.00059}_{-0.00053}$ |
| Flux ratio in the Kepler bandpass, $F_B/F_A$ | $0.8475^{+0.0110}_{-0.0076}$ | $0.3941^{+0.0011}_{-0.0010}$ | $0.01555^{+0.00010}_{-0.00006}$ |
| **Properties of the Stellar Binary Orbit** | | | |
| Period, $P$ (days) | $27.7958103^{+0.0000016}_{-0.0000015}$ | $20.733667^{+0.000012}_{-0.000012}$ | $41.079220^{+0.000078}_{-0.000077}$ |
| Semi-major axis length, $a$ (AU) | $0.22882^{+0.00019}_{-0.00018}$ | $0.17617^{+0.00028}_{-0.00029}$ | $0.22431^{+0.00035}_{-0.00034}$ |
| Eccentricity, $e$ | $0.52087^{+0.00052}_{-0.00055}$ | $0.1421^{+0.0014}_{-0.0014}$ | $0.15944^{+0.00061}_{-0.00062}$ |
| Eccentricity Times Sine of Arg. of Periapse, $e\sin(\omega)$ | $0.49377^{+0.00057}_{-0.00060}$ | $0.1418^{+0.0014}_{-0.0014}$ | $-0.15840^{+0.00063}_{-0.00061}$ |
| Eccentricity Times Cosine of Arg. of Periapse, $e\cos(\omega)$ | $0.165828^{+0.000065}_{-0.000061}$ | $0.0086412^{+0.0000031}_{-0.0000031}$ | $-0.0181481^{+0.0000045}_{-0.0000044}$ |
| Mean Longitude, $\lambda \equiv M + \omega$ (deg) | $300.1970^{+0.0099}_{-0.0105}$ | $89.1784^{+0.0011}_{-0.0012}$ | $92.3520^{+0.0011}_{-0.0011}$ |
| Inclination $i$ (deg) | $89.8584^{+0.0075}_{-0.0083}$ | $90.4238^{+0.0075}_{-0.0073}$ | $90.3401^{+0.0016}_{-0.0019}$ |

TABLE 1

Results of the photometric-dynamical model for Kepler-34 (KIC 8572936) and Kepler-35 (KIC 9837578). The orbital parameters listed are the osculating Jacobian parameters, *i.e.*, the instantaneous Keplerian elements for the listed epoch. In general, unlike the simple 2-body Keplerian case, the orbital elements are functions of time. In particular, the orbital period of Kepler-34's planet varies from 280-312 days on secular timescales; the median period is ~291 days. See the SI for details. For direct comparison, values[9] for Kepler-16 are listed.

**FIGURES**

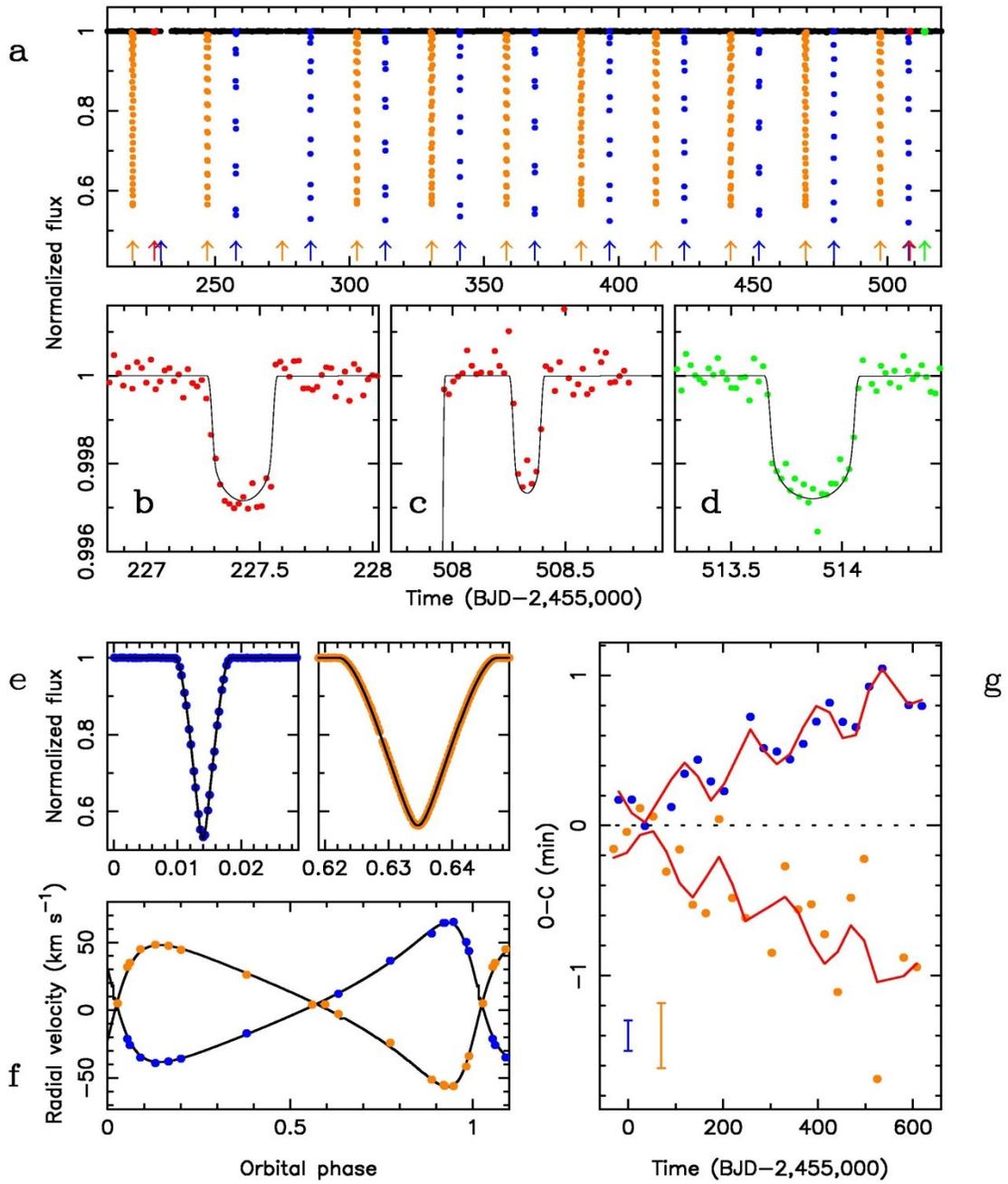

**Figure 1:** *Observations of Kepler-34.*

**Figure 1:** *Observations of Kepler-34.*
**(a)** A portion of the normalized light curve showing the relative brightness versus time (in units of barycentric Julian days BJD). Low-frequency variations and instrumental drifts have been removed (see SI). The blue points show the primary eclipses (star B eclipses star A), orange points show the secondary eclipses, red points show the primary transits (planet transits star A), and green shows the secondary transit. The times of each event are indicated by the arrows. Due to gaps in the observations, one primary and one secondary eclipse were missed.
**(b,c,d)** Close-up views of the three transit events. The solid curve is the photometric-dynamical model. Variations in transit widths are mainly due to differences in the transverse velocity of the stars during transit. The large drop before the transit in panel c is due to a primary eclipse.
**(e)** Close-up views of the phase-folded primary and secondary eclipses plotted versus orbital phase (time modulo the orbital period $P$, where $P=27.795794795$ d and the time of periastron is BJD 2,455,007.5190). Only *Kepler* Quarter 4 data are shown.
**(f)** Radial velocities of the primary star (blue dots), secondary (orange dots), and the model curve, versus orbital phase.
**(g)** Observed (O) minus computed (C) diagram showing the deviations between the measured eclipse times and those predicted assuming strict periodicity. Primary eclipses are shown as blue points, secondaries by orange points, and the corresponding models by the red curves. A period of 27.79578193 days and an epoch of BJD 2,454,979.72301 were used to compute the primary eclipse times, and a phase offset of 0.6206712 for the secondary eclipse times. The divergence indicates the primary and secondary periods are different. The two vertical bars in the lower left denote the median 1-sigma uncertainties of the primary and secondary eclipse times: 0.10 and 0.22 min.

**Figure 2:** *Observations of Kepler-35.*
The layout of this figure is similar to Fig. 1.
**(a)** A portion of the light curve for Kepler-35. Due to interruptions in the data acquisition, two primary and two secondary eclipses were not observed.
**(b,c,d,e)** Close-up views of the 4 transit events. The points in red denote primary transits, and the points in green denote a secondary transit. Note the differences in transit duration.
**(f)** Close-up views of the primary eclipses and secondary eclipses, plotted versus orbital phase where $P=20.733762175$ days and the time of periastron passage is BJD 2,455,007.3131. Only *Kepler* Quarter 4 data are shown (BJD 2,455,183 through 2,455,275).
**(g)** Radial velocities of the primary star (blue dots), secondary (orange dots) and model fit.
**(h)** Observed minus computed diagram, where a period of 20.73373997 days and an epoch of BJD 2,454,965.84579 were used to predict the primary eclipses, and a phase offset of 0.5055680 for the secondary eclipses. The two vertical bars in the upper left denote the median 1-sigma uncertainties of the primary and secondary eclipse times, 0.27 and 0.26 min, respectively.

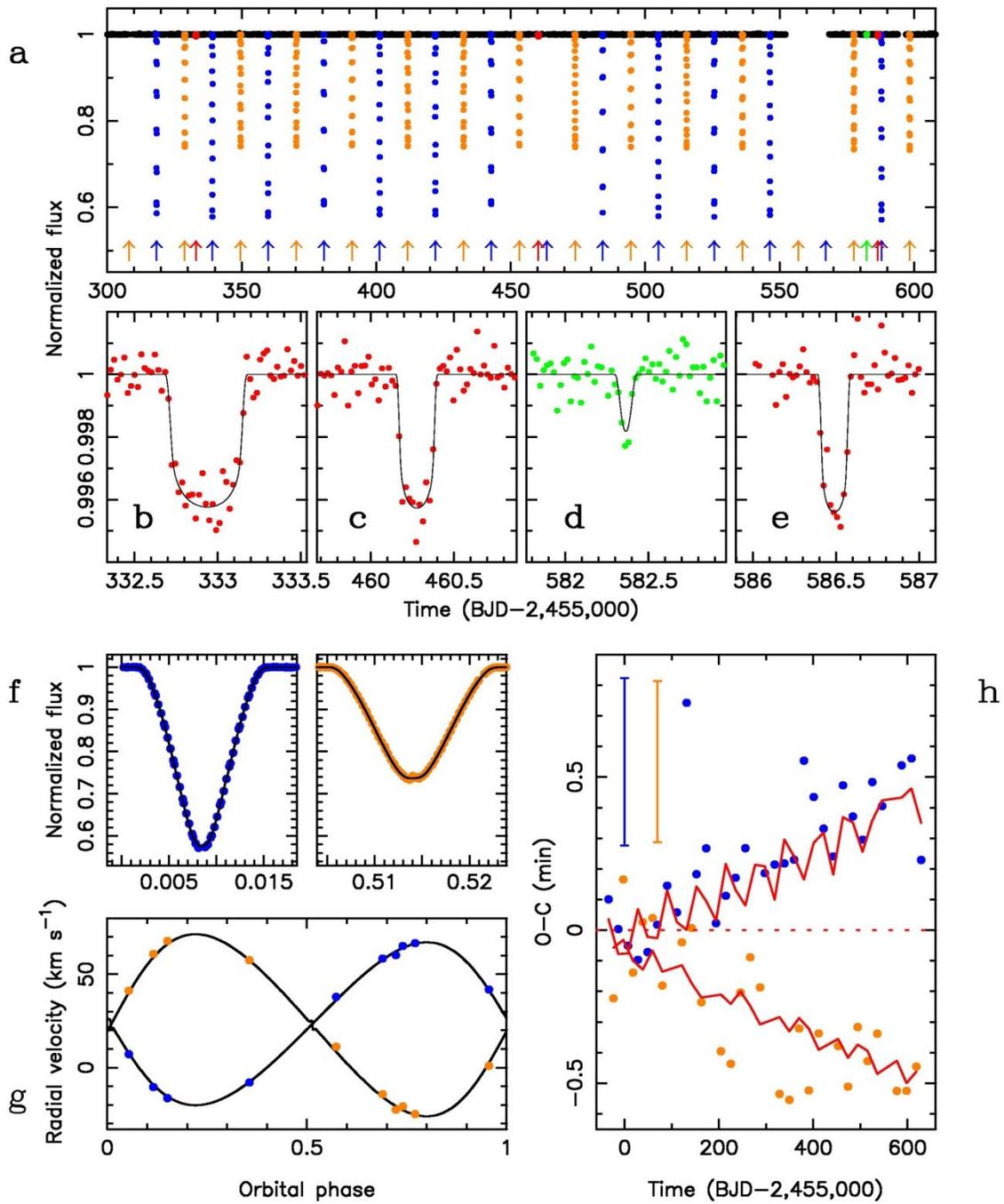

**Figure 2:** *Observations of Kepler-35.*

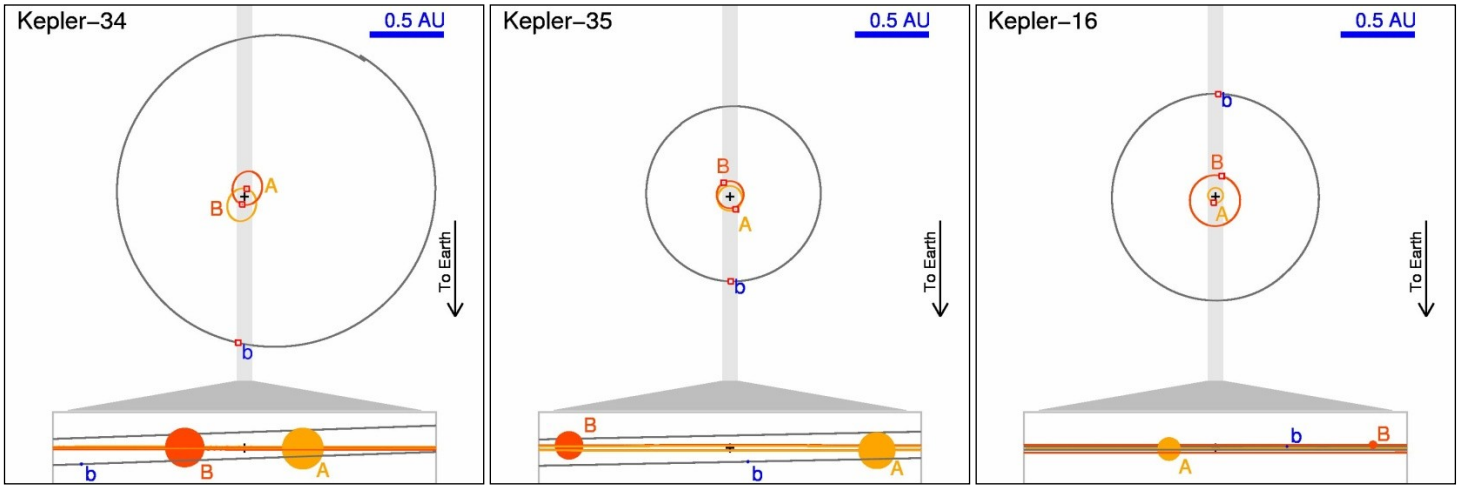

**Figure 3:** *Orbital configurations.*
(a) Left panel: A scale view of the orbits of the Kepler-34 system seen face-on and also as seen from Earth. In the face-on view, the stars and planet are too small to be seen relative to their orbit curves, and so are represented as dots and marked with symbols A, B, and b denoting the primary star, secondary star, and planet. This view is correct for a given epoch (BJD 2,455,507.50). Because of the dynamical interactions between the three bodies, this orbital configuration will evolve. For example, the orbits precess, and hence the orbits do not actually close.
The line-of-sight view shown in the box depicts the stars and planet with correct relative sizes and orientation. More importantly, the orbits and the orbital tilts are accurately portrayed, showing how transits do not necessarily occur at every conjunction.
**(b)** Centre panel: Same as for (a), but for Kepler-35 at epoch BJD 2,455,330.60. Note that the relative sizes of the bodies are drawn to scale for each panel (a,b,c) not just within a panel.
**(c)** Right panel: Same as for (a), but for Kepler-16 and at epoch BJD 2,455,213.0.

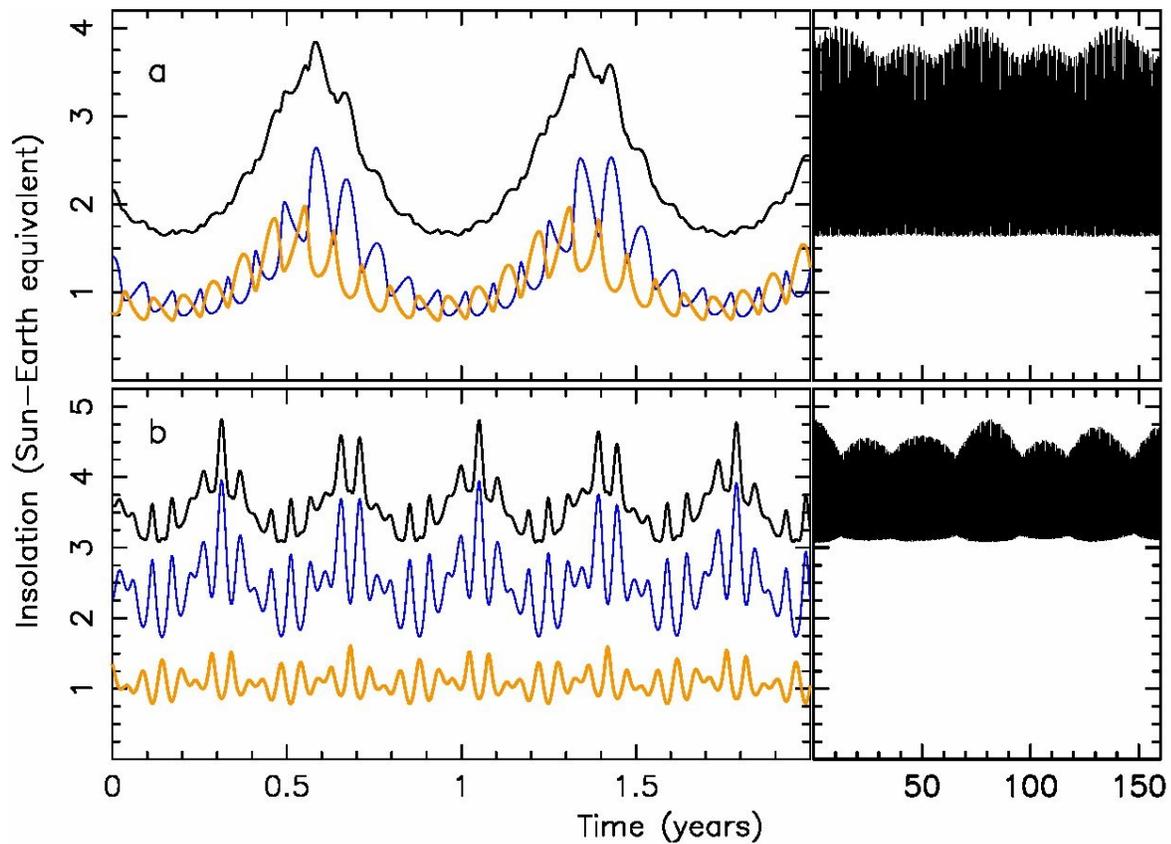

**Figure 4:** *Variations in insolation received by Kepler-34 and Kepler-35.*
**(a)** Top panels: The black curve shows the incident flux (insolation) received by Kepler-34 b from its two stars. The insolation is in units of the Solar constant *S* (solar flux received at a distance of 1 AU; *S*=1.0 for the Sun-Earth system). The contribution from star A is shown in blue and the contribution from star B in orange. The most rapid variations are caused by the orbital motion of the stars. The slower variations are due to the orbital motion of the planet. The right hand panel shows a longer timescale view of the insolation. The long-timescale quasi-periodicity is caused by the mutual precession of the orbits of the stars and planet, but is dominated by the precession of the planet.
**(b)** Lower panels: Same as (a) but for Kepler-35 b.



\def\lesssim{\lower2pt\hbox{$\buildrel {\scriptstyle <}
   \over {\scriptstyle\sim}$}}

\maketitle
\thispagestyle{empty}

\parindent 0pt


\parindent 0.39in

\vspace*{-0.1in}

\section{Alternate designations and summary of parameters}

Supplementary Tables S\ref{tab1} and S\ref{tab2}
give the alternate designations, coordinates, and magnitudes of Kepler-34
and Kepler-35.  These tables also summarize the system properties
as determined from spectroscopy (\S\ref{TODCOR}), eclipse timings
(\S\ref{ETV}), and the photometric-dynamical model (\S\ref{photodyn}).

\section{Optical imaging}

Blends of target stars with nearby stars on the sky can
be a serious problem with {\em Kepler} targets
since the contamination reduces the observed
eclipse and transit depths, which
might possibly lead to incorrect measurements of the component radii.
In order to assess the blends, 
we carried out imaging of the targets using
the Las Cumbres Observatory's  2.0 m Faulkes
Telescope North
at Haleakala, Hawaii.
Each image was combined from individual exposures taken at different
times of the night and on different nights, to average out the spider
pattern
and gain image depth while avoiding saturation. All images were in SDSS r
band, which is closest to the Kepler band among the broad band 
filters$^{14}$.
The pixel scale is 0.3 arcseconds per pixel, and the typical seeing
was 1.6 arcseconds full width at half maximum.

\begin{figure}
\centerline{\includegraphics[scale=0.89]{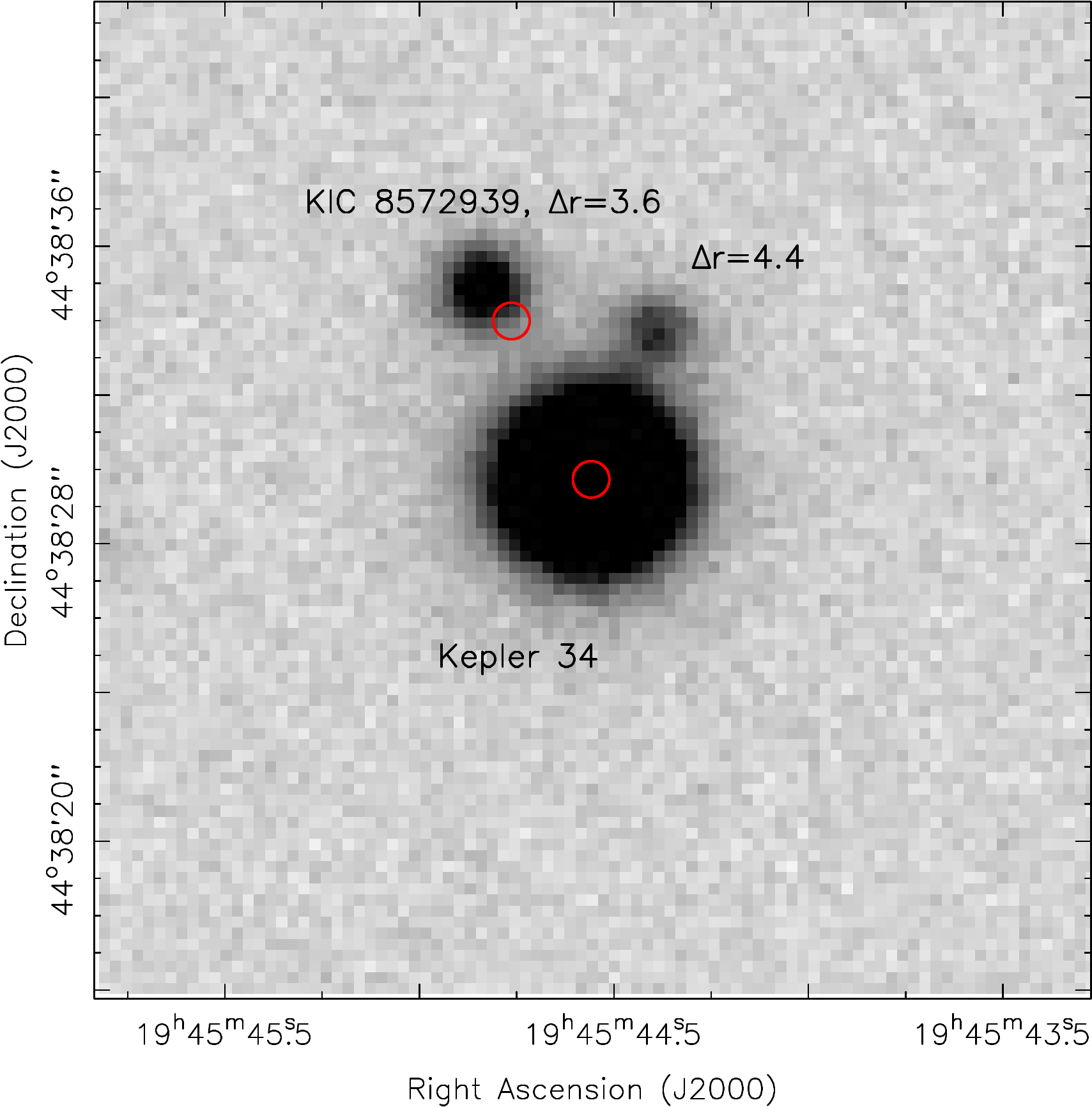}}
\caption{{\bf $\vert$ Image of Kepler-34.}
The $30^{\prime\prime} \times 30^{\prime\prime}$
region near Kepler-34 in the SDSS r filter.
There are three stars detected, and the two red circles (with
diameters of 1 arcsecond) mark the positions of the two
objects that appear in the KIC.  The actual
position of the brighter
neighbour star  
KIC 8572939 is about 1 arcsecond to the northeast.
That star is 3.6 mag
fainter in the SDSS r filter than Kepler-34, as estimated from
PSF photometry.
The faintest star does not appear in the KIC, and is about 4.4
mag fainter than Kepler-34 in SDSS r.
\label{2459_zoom}}
\end{figure}

Kepler-34 has a nearby star 4.5 arcsec to the northwest that is 4.4 mag
fainter in the SDSS r band (Supplementary Figure
S\ref{2459_zoom}).   This star does not appear in the Kepler
Input Catalog$^{14}$
(KIC), and as a result its flux contribution would not
be accounted for by the 
{\em Kepler} data analysis pipeline.  However, 
owing to its faintness, the additional contamination from 
this non-KIC star should be no
more than 1.7\%.  The star KIC 8572939, which is 3.6 mag fainter
than Kepler-34, is about 1 arcsecond northeast of its expected
position.

\begin{figure}
\centerline{\includegraphics[scale=0.91]{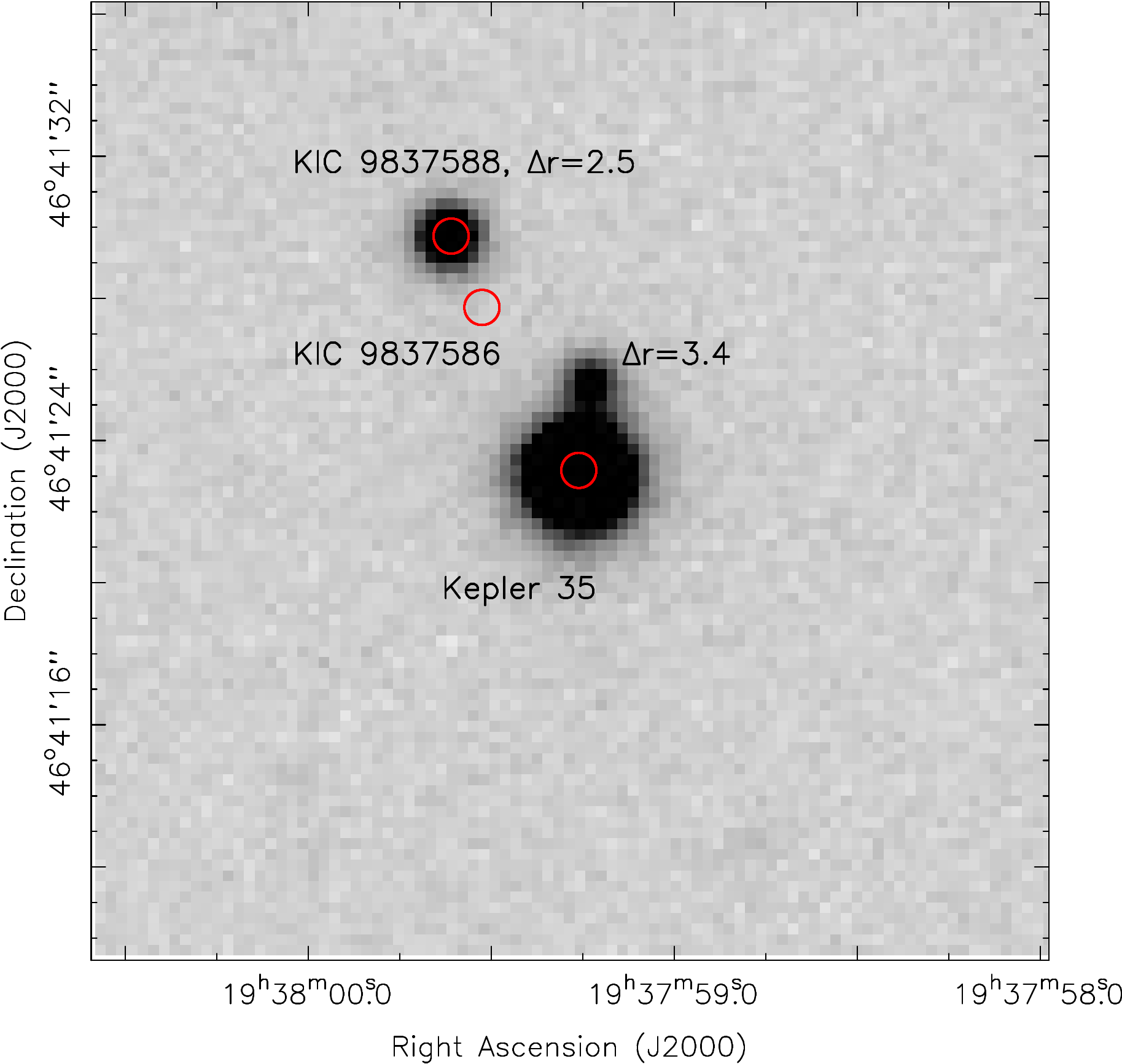}}
\caption{{\bf $\vert$ Image of Kepler-35.}
The $30^{\prime\prime} \times 30^{\prime\prime}$
region near Kepler-35 in the SDSS r filter.
There are three stars detected, and the three red circles (with
diameters of 1 arcsecond) mark the positions of the three
objects that appear in the KIC.  The position and magnitude
difference of KIC 9837588 (the northernmost star) is as
expected.  KIC 9837586, which is about 1.75 mag fainter
than Kepler-35,  should be between Kepler-35 and KIC 9837588, but
is apparently nowhere to be seen.  The fainter star just north
of Kepler-35 (which is not in the KIC)
is 3.4 mag fainter than Kepler-35, and is unlikely
to be KIC 9837586.
\label{2937_zoom}}
\end{figure}

Kepler-35 has a nearby star 2.5 arcsec to the north that is 3.4 mag
fainter in the SDSS r filter that does not appear in the
KIC (Supplementary Figure
S\ref{2937_zoom}). 
Assuming complete blending the additional
contamination is 4.2\%.  According to the KIC, Kepler-35 should have
two fainter neighbour star to the northeast.  However, only one of
them was detected.   KIC 9837588 is detected at its expected
position and at the expected brightness.  KIC 9837586,
which should be about 1.75 mag fainter than Kepler-35, is not
seen.  The anonymous star just north of Kepler-35 is not
likely to be KIC 9837586, as it
is about 1.7
mag fainter than the nominal brightness of KIC 9837586.

We conclude that the {\em Kepler} light curves both
Kepler-34 and Kepler-35 should only have modest contamination
($\lesssim\, 10\%$) due to nearby stars.   This excess light is
accounted for on a quarter-by-quarter basis in the
photometric-dynamical modelling discussed in \S\ref{photodyn}.

\section{Spectroscopic observations}

We observed Kepler-34 and Kepler-35 with the Hobby-Eberly Telescope (HET)
and the Harlan J. Smith 2.7 m Telescope (HJST) at McDonald Observatory
with the aim to help define the spectroscopic orbit of these two binary
systems. We used the High Resolution Spectrograph\cite{tull1998} 
(HRS)
at the
HET to collect 7 spectra for Kepler-34 in 2011 September and 4 spectra for
Kepler-35 in 2011 October. 
The HRS setup was equivalent to the instrumental
configuration we employ for most of our Kepler mission planet confirmation
work at the HET\cite{endl2011}. 
However, for these 2
targets we did not pass the starlight through the iodine cell.
Exposures times were 1800 s for Kepler-34 and 2700 s for
Kepler-35. During each visit to these  targets we also obtained
a spectrum of HD 182488, a RV standard star that we use to place
the RVs onto an absolute scale.  The images were reduced using
customized software.  The spectra have a resolving
power of $R=30,000$ and a wavelength coverage of
about 4800~\AA\ to 6800~\AA.

We used the Tull Coude Spectrograph\cite{tull1995} 
at the
HJST to observe Kepler-34 and Kepler-35. 
The Tull spectrograph covers the
entire optical spectrum at a resolving power of 
$R=60,000$. At each visit we
took three 1200 s exposures that we co-added to one 1 hour exposure. We
collected 14 1-h spectra for Kepler-34 over two observing runs in 
2011 September
and October. For Kepler-35 we obtained 5 1-h spectra in 2011
October.
Similar to the HET data we always observed the RV standard star HD 182488 in
conjunction with the targets.
The data were reduced and spectra were extracted using a reduction
pipeline developed for this instrument.

Kepler-35 was observed on 
2011 September 23-26
using the FIber-fed
Echelle Spectrograph (FIES) on the 2.5 m Nordic Optical Telescope (NOT) on
La Palma, Spain\cite{djupvik2010}. 
We used the medium resolution
fiber (1.3 arcsecond
projected diameter) with a resolving power of 
$R = 46,000$ giving
a wavelength coverage of about 
3600~\AA\ to  7400~\AA.  The total exposure times were 1 hour each.
The radial velocity standard star HD 182488 was also observed using
the same instrumental configuration.
The data were reduced and spectra were extracted using the FIES
pipeline\cite{buchhave2010}.

Spectra of Kepler-34 and Kepler-35 were obtained using the
10 m Keck 1 telescope   
and the HIRES spectrograph\cite{vogt1994}.  
The spectra   were collected using the standard planet
search setup and reduction\cite{marcy2008}. 
The resolving power is $R=60,000$ at
5500~\AA. Sky subtraction, using the ``C2 decker'' was implemented with
a slit that projects to $0.87 \times 14.0$ arcsec on the sky. 
The wavelength calibrations
were made for each night using
Thorium-Argon lamp spectra.

\begin{figure}
\centerline{\includegraphics[scale=0.7,angle=-90]{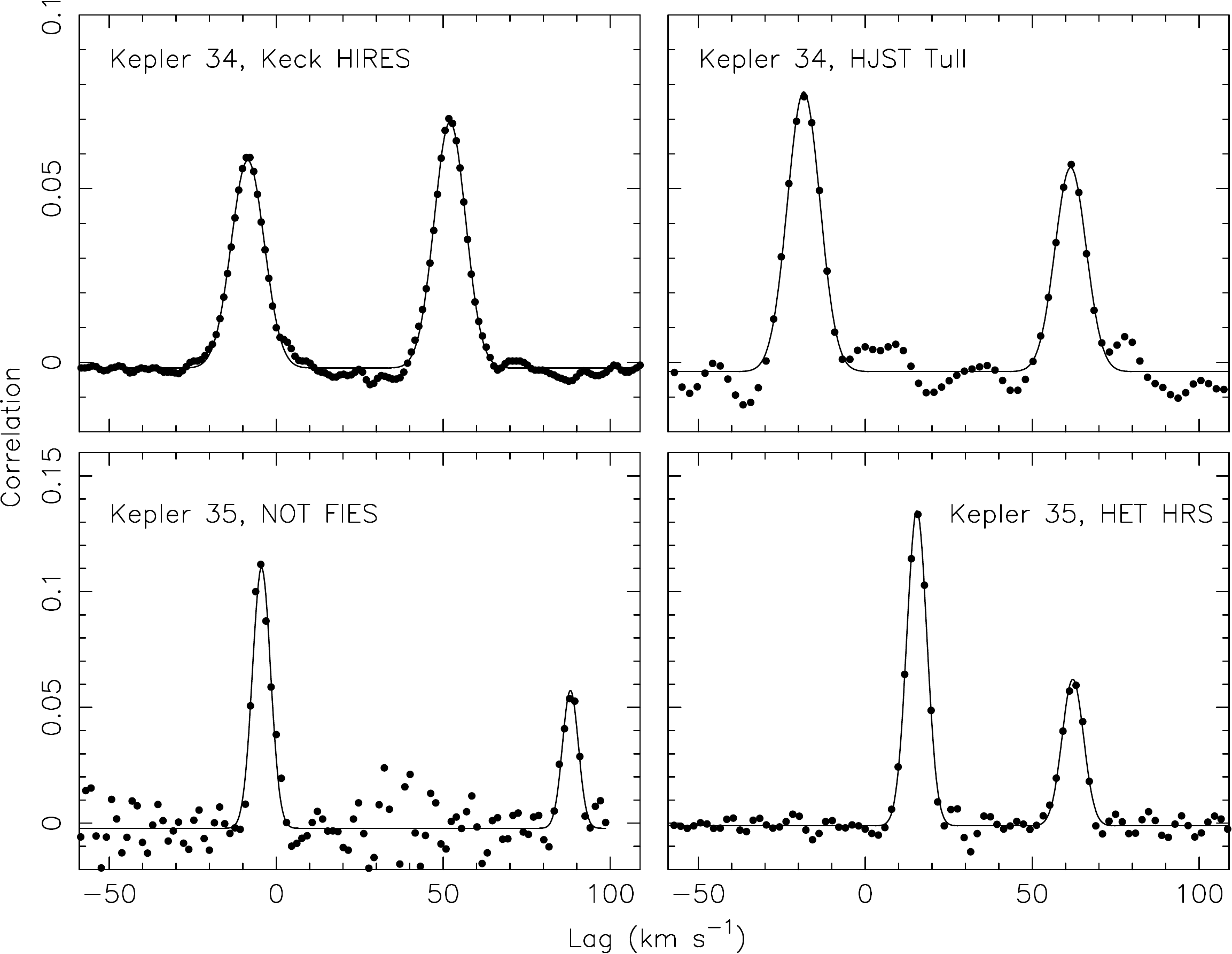}}
\caption{{\bf $\vert$ Broadening functions for Kepler-34
and Kepler-35.}
Four representative broadening functions
(filled circles)  for Kepler-34 and
Kepler-35 are shown.  The object and the
telescope and instrument is indicated
in each panel.  The solid lines are the best-fitting Gaussians.
\label{plotBF}}
\end{figure}

We used the ``broadening function'' technique\cite{rucinski1992} 
to measure the radial velocities.  
Observations of HD182488 (spectral type G8V)
were used
as the template star for each respective data set (HET, HJST,
FIES, and HIRES).  
The template radial velocity\cite{Nidever_2002} was assumed to be
$-21.508$
km s$^{-1}$.
The broadening functions (BFs) are essentially rotational
broadening kernels, where the centroid of the peak yields
the Doppler shift and where the width of the peak is
a measure of the rotational broadening.
Supplementary Figure S\ref{plotBF} shows four example BFs. In all cases,
the FWHM of the BF peaks were consistent with the instrumental
broadening, which indicates the rotational velocities are
not resolved.  Therefore, using the spectra with the highest
resolving power ($R=60,000$), we can place upper limits on
the projected rotational velocity of each star of 
$V_{\rm rot}\sin i\, \lesssim\, 5$ km s$^{-1}$.
The derived radial velocities for both stellar components
of Kepler-34 are given in Supplementary Table S\ref{RV2459}
and those for Kepler-35 in Supplementary Table S\ref{RV2937}.

\section{Spectroscopic parameters via TODCOR}\label{TODCOR}

Accurate temperatures and metallicity are essential for the
characterization of both the stars and the resulting planetary
environment, but the {\em Kepler}
photometric data do not provide strong
constraints on either parameter. The eclipses observed in the {\em
Kepler}
light curve yield the {\em ratio} $T_{{\rm eff},2}/T_{{\rm eff},1}$,
but only weakly constrain the absolute temperatures, and the
metallicity cannot be reliably determined photometrically. A
spectroscopic analysis can determine the effective temperature, surface
gravity, and metallicity, but all three parameters are highly
correlated and the results are unreliable in the absence of external
constraints. In transiting systems the mean stellar density can be
determined from the related light curve observable $a/R_*$ 
(see e.g.\ ref.\ \citen{soz07}),
effectively reducing the problem to a more manageable
$T_{\rm eff}-{\rm [m/H]}$ degeneracy. The same idea applies to
transiting circumbinary systems, although the photometric-dynamical model
employed here provides even stronger constraints -- a direct
determination of the stellar masses and radii, from which we
calculated the surface gravities. We then employed the two dimensional
cross-correlation routine TODCOR\cite{zuc94} and the
Harvard-Smithsonian Center for Astrophysics (CfA) library of synthetic
spectra to determine the effective temperatures of the binary members
and the system metallicity.

The CfA library consists of a grid of Kurucz model atmospheres\cite{kur05} 
calculated by John Laird for a linelist compiled by Jon
Morse. The spectra cover a wavelength range of $5050-5360$~\AA, and
have spacing of $250$~K in $T_{\rm eff}$ and $0.5$ dex in $\log g$ and
[m/H]. We cross-correlated the Keck/HIRES spectra with every pair of
templates spanning the range $T_{\rm eff}=[3000,7000]$, $\log
g=[3.5,5.0]$, ${\rm [m/H]}=[-1.0,+0.5]$, and recorded the mean peak
correlation coefficient at each grid point. Next, we interpolated to
the peak correlation value in each parameter (but fixed the surface
gravities to those found by the photometric-dynamical model) to determine the
best-fit parameters for the binary. Given the quality of the spectra,
we assigned internal errors of $100$~K in $T_{\rm eff}$ and $0.15$ dex
in [m/H] ($0.20$ dex for the weaker spectra of Kepler-35). However, as
mentioned above, the degeneracy between temperature and metallicity
could cause correlated errors beyond those quoted here. We explored
this by fixing the metallicity to the extremes of the $1$-$\sigma$
errors and assessing the resulting temperature offset. Incorporating
these correlated errors, we report the final parameters for Kepler-34: 
$T_{{\rm eff},1} = 5913 \pm 130$~K, $T_{{\rm eff},2} =
5867 \pm 130$~K, ${\rm [m/H]} = -0.07 \pm 0.15$; and for Kepler-35: 
$T_{{\rm eff},1} = 5606 \pm 150$~K, $T_{{\rm eff},2} =
5202 \pm 100$~K, ${\rm [m/H]} = -0.34 \pm 0.20$ dex.

Based on the scaling of the templates required to match the
observations and the flux ratio between the templates, TODCOR
provides a measurement of the ``luminosity ratio'' in the wavelength
range 5050-5360~\AA. For Kepler-34, we find
$ L_2/L_1 = 0.900 \pm 0.005$ and
for Kepler-35, we find 
$L_2/L_1 = 0.377 \pm 0.015$.

\section{Stellar rotation, gyrochronology, and tidal synchronisation}

Another relevant property that can be estimated is the
rotation period. 
For Kepler-34, outside of the eclipses the light curve exhibits
quasiperiodic variations with a peak-to-peak amplitude of about 0.06\%.
A power spectrum reveals a complex pattern of peaks, with most of the
power at periods of 15-18 days. The autocorrelation function also has
a strong, broad peak at 16 days. We interpret the periodicity as the
effect of starspots being carried around by stellar rotation. We
cannot say if one star is producing most of the observed variability,
or if it is a superposition of comparable signals from both stars, but
as the stars are similar in most respects it seems reasonable that
they both have a rotation period in the neighbourhood of 15-18 days.
Using the stellar radii in Supplementary Table S\ref{tab2}, this gives
a projected rotational velocity of $V\sin i\approx 3$ to 4 km s$^{-1}$,
consistent with the observed upper limit of $\approx 5$ km s$^{-1}$
(assuming the angular momentum vector of the stellar rotation is aligned
with the angular momentum vector of the orbit).

Sun-like stars are rapid rotators when they are young, and spin down
as they age, with an approximate dependence $P_{\rm rot} \propto
t^{1/2}$ and a secondary dependence on stellar mass or spectral type.
Therefore the measured rotation period and mass can be used to
determine a ``gyrochronological'' age for the stars. Since the measured
rotation period  is shorter than the Sun's rotation period of 25.4
days, one would expect these stars to be younger than the Sun's
main-sequence age of 4.5 Gyr. For a more accurate comparison we used
an age-mass-period model\cite{schlaufman2010}
which gives a
gyrochronological age of 2.0-2.9 Gyr for the primary star and 1.9-2.7
Gyr for the secondary star in Kepler-34
(with the uncertainty range representing
only the uncertainty in the rotation period).

There is a dissonance between the gyrochronological age of
2-3 Gyr and the age of 5-6 Gyr that we determine from
comparison of the spectroscopic properties with theoretical
evolutionary models (\S\ref{evol})).
There is reason to suspect the gyrochronological
age, because the tidal forces in this close binary have probably had
enough time to alter the spin rates by a significant degree.

Tidal torques act to synchronise the rotation and orbital periods, and
circularise the orbit, with circularisation taking longer than
synchronisation. Before circularisation is achieved, most tidal
theories predict that the stars should become ``pseudosynchronised'',
reaching a spin period for which there is a vanishing tidal torque
when averaged over an orbit. In the specific tidal model of 
ref.\ \citen{hut1981}
the pseudosynchronous period would be 9.24 days for a binary with the
observed eccentricity of Kepler-34, which is shorter than the observed
rotation period.  Apparently the stars have not achieved
pseudosynchronisation, although it is still certainly possible that
the spin rates have been significantly altered by tides.

Finally, we examined the Ca II H\&K region of the Keck spectra of
Kepler-34 for signs of chromospheric activity. Unfortunately, the 
signal-to-noise is only about
5 for this region.  Qualitatively, there are no signs of exceptional 
activity in the Ca II H\&K
lines.

For Kepler-35, the light curve is somewhat noisy 
(white noise $rms \sim 590$ ppm) 
due to the relative faintness of the star ($Kp=15.7$), 
but a modulation at roughly 20 days is clearly visible by eye. 
The power spectrum is clean and shows a strong spike at 
$20.8 \pm 0.1$ days, and the autocorrelation function shows a 
broad peak at 21 days. This periodicity agrees perfectly with 
the binary orbital period ($P=20.734$ d). In addition, the 
shape of the modulation is fairly sinusoidal, not ``W''-shaped 
that is often associated with starspot modulations. Thus we 
conclude that the photometric modulation is not related to 
stellar activity (i.e., starspots), and that we cannot measure 
the rotation period of the star via the photometry. 
However, the lack of any measurable stellar activity does 
suggest an old age for the star, consistent with the age 
derived in \S\ref{evol} via stellar evolution models. An interpretation 
of the orbital period modulation, Doppler beaming, is presented 
in \S\ref{beam}.

\section{Light curve preparation and detrending}

For the binary star and planet modelling 
we use the basic ``raw'' or  ``PA'' photometry provided by the
{\em Kepler} pipeline and available at the MAST archive.
{\em Kepler} light curves often show instrumental trends, so 
we did further processing to
detrend the data.  
In general, each quarter of data must be detrended separately, since
after the spacecraft makes its quarterly rolls to align its solar
panels to the Sun
the target star will appear on a
different detector module.
The software used to measure eclipse times and 
the photometric-dynamical model discussed below use their
own local detrending algorithms.  
Separate globally
detrended light curves were also made for use in 
Figures 1 and 2, and also for independent light curve modelling checks. 
Here, the basic detrending process is an iterative clipping
technique. Detrending is complicated by the presence of eclipses in the
light curve which must be removed before detrending can be done. The
basic process
for this is the data is fit to a Legendre
polynomial of order $k$, where $k$ is typically very high (60-200). Then
sigma-clipping is done so any points $3\sigma$ above or below the fit
are discarded. Then the fit is recalculated, and again
sigma-clipped. This is repeated until all eclipses or other
discontinuities, such as those caused by cosmic rays, are removed,
allowing the final fit to be subtracted from the original data,
providing a detrended light curve.

The PA and detrended light
curves  for Kepler-34 and Kepler-35
are shown in Supplementary Figures S\ref{plotraw2459} and
S\ref{plotraw2937}, respectively.
In some cases an eclipse was interrupted by a gap in the observing.
Since incomplete coverage may introduce errors
in the detrending, we excluded partially
observed events entirely.

\section{Doppler beaming}\label{beam}

The 
{\em Kepler} precise light curves can reveal the beaming effect (aka
Doppler boosting) of short-period binaries, an effect that causes the
stellar intensity to modulate because of the stellar radial-velocity
periodic motion\cite{loeb2003,zucker2007}. 
The amplitude of the Doppler
beaming is on the order of $4V_{rel}/c$, where $V_{rel}$ is the
radial velocity of the source relative to the observer and $c$ is the
speed of light\cite{rybicki1979}. 
Usually, the beaming 
modulation appears together with two well known effects, the
ellipsoidal\cite{mazeh2008} 
and the reflection\cite{for2010}
effects.

To derive the beaming effect of Kepler-35 due to the stellar orbits,
we performed a long-term
detrending of the light curve with a cosine filter\cite{mazeh2010},
ignored the eclipses, and then fitted the detrended data with a model
that included the ellipsoidal, beaming and reflection effects
(hereafter the BEER model, following 
ref.\ \citen{faigler2011}). 
We
approximated the beaming and the ellipsoidal modulations by pure 
sine/cosine functions, using mid-primary eclipse timing and the period 
derived in this work.
The beaming effect was represented by a sine
function with the orbital period, and the ellipsoidal effect by a
cosine function with half the orbital period. The reflection was
approximated by the Lambert law\cite{demory2011}.

Supplementary
Figure S\ref{model_fit} shows the best-fit BEER model and
Supplementary 
Table S\ref{table_coeff} lists the resulting amplitudes. Only the
beaming effect is highly significant,
with an amplitude of $214\pm 5.7$ ppm.
This is not surprising, as
the beaming effect is expected to be
much larger than the other two modulations when the binary period is
longer than 10 days\cite{loeb2003,zucker2007}. 
 When we adopt the binary-orbit elements from the
photometric and radial-velocity solution we derive an amplitude of
$230\pm6$ ppm, not very different from the amplitude of the sine
function.

The observed beaming modulation is the sum of the
effect of the primary and that of the secondary\cite{zucker2007},
which depend on the stellar temperatures, fluxes and masses. If we
know the temperatures and the radial-velocity amplitudes of the two
stars, we can in principle derive the flux ratio from the amplitude
of the observed beaming effect. In our case, we derive a flux ratio
of $\sim0.4$, consistent with the value derived from the eclipse
analysis and from the spectra.

\section{Measurements of eclipse times}\label{ETV}

The times of mideclipse for all primary and secondary
events in Kepler-34 and Kepler-35 were measured in a
manner similar to that described in ref.\ \citen{steffen2011}.
Briefly, the times of primary eclipse and the times
of secondary eclipse are measured separately for each source.
Given an initial linear ephemeris and an
estimate of the eclipse width, the data around the eclipses
were isolated, and locally detrended with a cubic polynomial
(the eclipses were masked out of the fit).  The detrended
data were then folded on the linear ephemeris, and a cubic
Hermite spline fit was used to make an eclipse template.
The template was then iteratively correlated with each eclipse to
produce a measurement of the eclipse time. This time was then corrected
to account for the Long Cadence 29.4244 minute bin size,
which otherwise could induce an alias periodicity.

Supplementary Figure S\ref{plotprof} shows the templates and
folded data for Kepler-34 and Kepler-35.  Generally, the
template profiles are an excellent match to the folded
data.  There are few points near mideclipse (both primary
and secondary) in both Kepler-34 and Kepler-35 that
are much brighter than other nearby points.  These
anomalous points, which are somewhat common in {\em Kepler}
light curves of deeply eclipsing binaries, are the result of
undesirable behavior in the cosmic ray detection routines
used in the data analysis pipeline.   The anomalous events are
believed to
happen for these types of eclipses because
(i) the size of the
windows used to detrend the data
in order to identify impulsive outliers is comparable to the
eclipse width; 
(ii) small changes in pointing can result in
significant changes in pixel flux near the core of star images;
and (iii) the stellar intensity is rapidly changing owing to the
eclipse.  These three conditions can sometimes lead
the routines to flag good data at mideclipse as a negative
outlier and incorrectly apply a positive cosmic ray correction
(note that
cosmic rays are flagged at the pixel level before the
flux time series is constructed).
The cosmic ray detection routines are not restricted
to identify only
positive outliers because there are known
sources of impulsive negative
outliers.
These anomalous events 
in the Kepler-34 and Kepler-35 were identified, and the uncertainties on the
fluxes are increased by a factor of 100, effectively clipping them from the
light curves.

The times of mideclipse for both primary and secondary
eclipses for Kepler-34 and Kepler-35 are given in
Supplementary Tables S\ref{ETV34} and S\ref{ETV35},
respectively.  The cycle numbers for the secondary
are not exactly half integers owing to the eccentric
orbits.
A linear ephemeris was fit to each set,
resulting in the Observed minus Computed (O--C) diagrams
shown in Supplementary Figure S\ref{plotoc}.  The curves
are generally flat, although the O--C plot for the Kepler-34
primary eclipse shows modest power at a period of 137 days,
which is roughly one half of the period of the planet at the
current epoch.
The best-fitting ephemerides for each set are

\begin{tabular}{rcr@{\,$\pm$\,}ll}
$P_A$ & = & $27.7958070$ & $0.0000023$      & Kepler-34 primary  \\
$P_B$ & = & $27.7957502$ & $0.0000065$      & Kepler-34 secondary  \\
$T_0(A)$ & = & $54979.72308$ &  $0.000036$ & Kepler-34 primary    \\
\vspace{1em}
$T_0(B)$ & = & $54969.17926$ &  $0.000085$ & Kepler-34 secondary    \\
$P_A$ & = & $20.7337496$ & $0.0000039$    & Kepler-35 primary \\
$P_B$ & = & $20.7337277$ & $0.0000040$    & Kepler-35 secondary \\
$T_0(A)$ &=& $54965.84580$ &  $0.000034$ & Kepler-35 primary \\
$T_0(B)$ &=& $54976.32812$ &  $0.000033$ & Kepler-35 secondary \\
\end{tabular}

\noindent where the periods are in days and the
reference times are in units of BJD - 2,400,000.  The primary
and secondary periods in Kepler-34 differ by $4.91\pm 0.59$ seconds.
The corresponding period difference for Kepler-35 is $1.89\pm 0.48$
seconds.

Given the precision that we can measure eclipse times, and the closeness
of these circumbinary gas-giant planets to their habitable zones, it is 
interesting to consider the presence of moons around these planets. 
Unfortunately,
the photometric signal for a Galilean-size or even Earth-size moon
is too small to measure in individual transits for these faint systems
(Kepler magnitudes of 14.9 and 15.7 mag).
Timing variations are another potential way to detect moons. However,
unlike the transit timing variations in single-star systems, here the
dynamical signatures are in the eclipse timings of the stars, not the
planets. The presence of a moon orbiting a circumbinary planet will
have no measurable effect on the stellar eclipse timing variations.
Meanwhile, the 
times of the planet transits can vary by several {\em days} without the 
presence of a moon. For Kepler-35, the time intervals between primary 
transits is 127.3 d, 122.1 d, and 126.2 d. Like the transit durations, 
the transit intervals vary due to the orbital motion of the stars: the
location of the star in its orbit at the time of conjunction can vary from 
transit to transit. By comparison, the shift in transit times due to the 
presence of a moon is only of order seconds to tens of seconds, making 
such a detection
infeasible,
especially with
the Long Cadence data (29.4 minute sampling)
obtained for these systems.

\section{Photometric-dynamical model}\label{photodyn}

The photometric-dynamical model was used in the Kepler-16
and KOI-126 investigations$^{9,15}$
and for completeness we
repeat a full description of the model and its
application to Kepler-34 and Kepler-34 here.

\subsection{Description of the model:}

The ``photometric-dynamical model'' refers to the model$^{15}$
that was used to fit
the {\it Kepler} photometry and the radial-velocity data for both 
Kepler-34
and Kepler-35.  The underlying model was a gravitational three-body
integration.  This integration utilized a hierarchical (or Jacobian)
coordinate system. In this system, ${\bf r_1}$ is the position of Star B
relative to Star A, and ${ \bf r_2}$ is the position of Planet b relative to
the centre of mass of the stellar binary (AB).  The computations are
performed in a Cartesian system, although it is convenient to express ${\bf
r_1}$ and ${\bf r_2}$ and their time derivatives in terms of osculating
Keplerian orbital elements: instantaneous period, eccentricity, argument of
pericentre, inclination, longitude of the ascending node, and mean anomaly:
$P_{1,2}$, $e_{1,2}$, $i_{1,2}$, $\omega_{1,2}$, $\Omega_{1,2}$, $M_{1,2}$,
respectively.

The accelerations of the three bodies are determined from Newton's equations
of motion, which depend on ${\bf r_1}$, ${\bf r_2}$ and the 
masses\cite{soderhjelm1984,mardling2002}.
An additional term is added to the acceleration of ${\bf r}_1$ to
take into account the leading order post-Newtonian potential of the stellar
binary\cite{soffel1989}. 
The computation is performed in units such that Newton's
gravitational constant $G \equiv 1$.  For the purpose of reporting the
masses and radii in Solar units, we assumed $G M_{\rm Sun} = 2.959122 \times
10^{-4}$ AU$^{3}$ day$^{-2}$ and $R_\odot = 0.00465116$ AU.  For the planet,
we report in Jupiter units with $M_{\rm Jupiter}/M_{\odot} = 0.000954638$
and $R_{\rm Jupiter}/R_{\odot} =  0.102792236$.

We used a Bulirsch-Stoer algorithm\cite{press2007} 
to integrate the coupled
first-order differential equations for $\dot{\bf r}_{1,2}$ and ${\bf
r}_{1,2}$.  For comparison between the model calculations and the observed
data at a given time, the Jacobian coordinates (${\bf r_1}$ and ${\bf r_2}$
and their time derivatives) are transformed into the
ordinary spatial coordinates of the three bodies relative to the barycentre
(the centre of mass of the entire three-body system). The instantaneous
positions of the three bodies were then projected to the location of the
barycentric plane (the plane that contains the barycentre and is
perpendicular to the line of sight), correcting for the delay resulting from
the finite speed of light.

The radial velocities of the stars were computed from the time derivative of
the position along the line of sight. The computed flux was the sum of the
fluxes assigned to Star A, Star B, and a constant source of ``third light,''
minus any missing flux due to eclipses.  The third light was specified for
each of the eight available quarters of 
{\em Kepler} data so as to account for
variable aperture size and spacecraft orientation. The loss of light due to
eclipses was calculated as follows. All objects were assumed to be
spherical. The sum of the fluxes of Star A and Star B was normalized to
unity and the flux of Star B was specified relative to that of Star A. The
radial brightness profiles of Star A and Star B were modelled with a
quadratic limb-darkening law, i.e., $I(r)/I(0) = 1-u_1 (1-\sqrt{1-r^2})-u_2
(1-\sqrt{1-r^2})^2$ where $r$ is the projected distance from the centre of a
given star, normalized to its radius, and $u_1$ and $u_2$ are the two
quadratic limb-darkening parameters\cite{limb}.

\subsection{{Specification of parameters:}}
The model has 35 adjustable parameters for each system. Three are mass
parameters ($\mu_A \equiv G M_A$, $\mu_B$, $\mu_C$). Six parameters are the
osculating orbital elements of planet b's orbit around the stellar binary AB
at a particular reference epoch $t_0$ ($P_2$, $e_2 \sin \omega_2$, $e_2 \cos
\omega_2$, $i_2$, $\lambda_2 \equiv \omega_2+M_2$, $\Omega_2$). The
reference epoch was selected to be near the time of a primary eclipse in
both systems and is listed in Table 1.  Five parameters are the osculating
orbital elements of the stellar binary at $t_0$ ($P_1$, $e_1$, $\omega_1$,
$i_1$,$M_1$).  The longitude of the ascending node of the stellar binary
relative to celestial North is unconstrained. For simplicity, it was held
fixed at $\Omega_1 = 0^\circ$, and hence $\Omega_2$ should be regarded as
the angle between the longitude of nodes of Planet b's circumbinary orbit,
and the longitude of nodes of the stellar binary orbit.

Three more parameters involve the radii of the bodies: the radius of Planet
b ($R_b$) and the relative radii of Star A and Star B ($R_A/R_b$,
$R_B/R_b$).  Five more parameters, related to the brightness profiles of the
stars, are the ratio of {\it Kepler}-bandpass fluxes of the stars
($F_B/F_A$) and the four limb-darkening coefficients of Star A and Star B
($u_1$, $u_2$ for each star). Eight additional parameters specify the
constant third light over a given 
{\em Kepler} quarter.  Another three parameters
were constant offsets representing the difference between the three
spectrographs' (TRES, HIRES, and McDonald with 
Kepler-34, and HET, HIRES,
and FIES with Kepler-35) radial-velocity scales  and the true line-of-sight
relative velocity of the barycentres of the Solar system and of 
Kepler-34 or
Kepler-35; this is needed because the radial-velocity variations are known
more precisely than the overall radial-velocity scale. Finally, there were
three parameters describing the photometric and radial velocity noise
profiles, both assumed to be white and Gaussian-distributed ($\sigma_A$,
$\sigma_B$, and $\sigma_{\rm phot}$, described further below).  

\subsection{Photometric data selection:}

The {\it Kepler} photometric data utilized in the final posterior
determination is a subset of the total data available for Q1
through Q8.  In particular, only the data within two durations of a
given eclipse (stellar or planetary) were retained.  Each continuous
segment about an eclipse was divided by a linear correction with time
to account for systematic trends on long timescales common in {\it
Kepler} data.  This linear correction was determined by fitting the
data outside of eclipse with a robust fitting algorithm.

\subsection{{Best-fitting model and residuals:}}
The likelihood ${\cal L}$ of a given set of parameters was taken to be the
product of likelihoods based on the photometric and radial-velocity data,
each of which was taken to be proportional to $\exp(-\chi^2/2)$ with the
usual definition of $\chi^2$, viz.,
\begin{eqnarray}
        {\cal L} &\propto& \left(2 \pi \sigma_{\rm
phot}^2\right)^{-\frac{N_{\rm phot.}}{2}} \exp \left(-\sum_i \frac{\Delta
F_i^2}{2 \sigma_{\rm phot}^2} \right) \times \\ \nonumber \\
                && \left(2 \pi \sigma_{\rm A}^2 \sigma_{\rm
B}^2\right)^{-\frac{N_{\rm RV}}{2}}  \exp \left(-\sum_j \frac{\Delta {\rm
RV_A}^2_j}{2 \sigma_A^2 \sigma_{A, j}^2}\right) \times \exp \left(-\sum_j
\frac{\Delta {\rm RV_B}^2_j}{2 \sigma_B^2 \sigma_{B, j}^2}\right) \nonumber
\end{eqnarray}
where $\Delta F_i$ is the $i$th photometric data residual, $\Delta {\rm
RV_{(A, B)}}_j$ and $\sigma_{(A,B)j}$ is the $j$th Star A or Star B radial
velocity residual and velocity uncertainty (see Supplementary Table 
S\ref{RV2459}).  The free
parameters $\sigma_A$, $\sigma_B$, and $\sigma_{\rm phot}$ specify the noise
profile of the RV data and photometric data.  The RV noise scaling factors
$\sigma_A$ and $\sigma_B$ were applied independently to velocities for Star
A and Star B, respectively.  These scaling factors account for systematic
sources of noise not captured in fits to the broadening functions and may
include night-to-night stability errors.  As may be expected, the RV noise
scaling factors were greater than one for both stars in both systems.
The increase in the RV errors results in larger
errors for the remaining parameters.

The best-fitting model was obtained by maximizing the likelihood.  
Supplementary Figures S\ref{output857} and S\ref{output983}
show the photometric data, the best-fitting model, and
the differences between the data and the best-fitting model for
Kepler-34 and Kepler-35, respectively.

\subsection{{Parameter estimation:}}
After finding the best-fitting model, we explored the parameter space and
estimated the posterior parameter distribution with a Differential Evolution
Markov Chain Monte Carlo (DE-MCMC) algorithm\cite{braak2006}. 
In this algorithm,
a large population of independent Markov chains are calculated in parallel.
As in a traditional MCMC, links are added to each chain in the population by
proposing parameter jumps, and then accepting or denying a jump from the
current state according to the Metropolis-Hastings criterion, using the
likelihood function given in Section 2.3 of this supplement. What is
different from a traditional MCMC is the manner in which jump sizes and
directions are chosen for the proposals.  A population member's individual
parameter jump vector at step $i+1$ is calculated by selecting two randomly
chosen population members (not including itself), and then forming the
difference vector between their parameter states at step $i$ and scaling by
a factor $\Gamma$.  This is the Differential Evolution component of the
algorithm. The factor $\Gamma$ is adjusted such that the fraction of 
accepted jumps, averaged over the whole population, is approximately 25\%.

We generated a population of 128 chains and evolved through approximately
1500 generations.  
The initial parameter states of the 128
chains were randomly selected from an over-dispersed region in parameter
space bounding the final posterior distribution.  The first 30\% of the
links in each individual Markov chain were clipped, and the resulting chains
were concatenated to form a single Markov chain, after having confirmed that
each chain had converged according to the standard criteria.  In particular,
we report that the Gelman-Rubin statistic was less than $1.2$ for all
parameters. The values reported in Table 1 were found by computing the 50\%
level of the cumulative distribution of the marginalised posterior for each
parameter. The quoted uncertainty interval encloses 68\% of the integrated
probability around the median.  Supplementary Figures 
S\ref{mcmc857} and S\ref{mcmc983}
show many
of the two-parameter joint distributions for each system, highlighting many
of the strongest correlations that are seen. 

\section{Comparison to stellar evolution models}\label{evol}

The very precise stellar mass and radius determinations for 
Kepler-34
($\sigma_M/M$ and $\sigma_R/R$ less than 0.3\%) and Kepler-35
($\sigma_M/M < 0.6\%$, $\sigma_R/R < 0.3\%$), along with our
measurement of the effective temperature
and metallicity of the stars, offers the opportunity to compare
against models of stellar evolution, which in turn yields age
estimates for the two systems. The comparison for Kepler-34
is shown in
Supplementary
Figure S\ref{KOI2459iso}, 
where the left panel displays evolutionary tracks$^{16}$
(solid
lines) from the series  calculated for the exact masses
measured for the primary and secondary stars. The tracks are computed
for the metallicity that best fits the measured temperatures, which is
${\rm [Fe/H]} = -0.02$. 
This composition is consistent with the metallicity of
${\rm [m/H]} = -0.07 \pm 0.15$ determined spectroscopically. The temperature
difference from spectroscopy is in excellent agreement with that
predicted by the models, which implies consistency with the
measured mass ratio.  The dotted lines in the figure represent two
isochrones for the best-fit metallicity and ages of 5 Gyr and 6 Gyr,
which bracket the measurements. According to these models, the system
is therefore slightly older than the Sun.  On the right-hand side of
Supplementary
Figure S\ref{KOI2459iso} 
the measured radii and temperatures of the two stars are
shown separately as a function of mass. The same two isochrones are
plotted for reference, showing the good agreement with theory.

A similar diagram for Kepler-35 is shown in 
Supplementary Figure S\ref{KOI2937iso}. 
In this case the
best-fit metallicity is ${\rm [Fe/H]} = -0.13$, also consistent with the
spectroscopic determination of 
${\rm [m/H]} = -0.34 \pm 0.20$. Once again
there is agreement between the temperature difference measured
spectroscopically and that inferred using models for the measured
masses.  The age of the system is more poorly determined than in
Kepler-34, but appears to be considerably older.  The dotted lines in
the figure correspond to isochrones for the best-fit metallicity and
ages of 8 Gyr to 12 Gyr, which we consider to be a very conservative
range for this system. The measurements in the mass-temperature
diagram on the right-hand side of 
Supplementary Figure S\ref{KOI2937iso} 
show good agreement with
theory, but the measured radii suggest a somewhat steeper slope in the
mass-radius plane than indicated by the isochrones. The source of this
discrepancy is unclear. The system would benefit from additional
spectroscopic observations to reach definitive conclusions.

The distances can be estimated to Kepler-34 and Kepler-35
using the parameters in Supplementary Tables S\ref{tab1}
and S\ref{tab2}.  The absolute magnitudes of the stars
in a given filter bandpass (in particular the
2MASS $J$ filter)
can be computed given their radii, temperatures,
and gravities using filter-integrated fluxes computed
from detailed model atmospheres\cite{allard}.
The apparent magnitude of the source $J$ and
$J$-band interstellar extinction then lead to the distance.
We find $d=1499\pm 33$ pc for Kepler-34
and $d=1645\pm 43$ pc for Kepler-35.

\section{Forward integration and stability}\label{stable}

\subsection{{Secular variations in orbital parameters:}}
Supplementary
Figures S\ref{els857} and S\ref{els983}
shows the time variation of selected
orbital elements of the planet's orbit in both systems
over 100 years, relative to the
invariable plane (the plane perpendicular to the total angular momentum of
the system). The positions and velocities of the masses were recorded with a
time sampling of 5~days. The slow (secular) variations in the orbital
elements occur on a timescale of approximately 30 to 70 years for Kepler-34
depending on the orbital element 
and 10 to 30 
years for Kepler-35.  

\subsection{{Long-term stability:}}
According to the approximate criteria for
dynamical stability$^{17}$,
the nominal
models for Kepler-34 and Kepler-35 systems are sufficiently widely
spaced to be dynamically stable. Nevertheless, we performed direct
$N$-body integrations to test the stability of both systems. For the
nominal solutions (Table 1), we integrated for ten million years using
the conservative Burlisch-Stoer integrator\cite{press1992}
in
Mercury v6.2 
(ref.\ \citen{mercury})
and found no
indications of instability. In addition, we tested one thousand
systems with masses and orbital parameters drawn from the posterior
distribution according to the DEMCMC algorithm described in SI 
Sec.\ \ref{photodyn}.
For each of these, we integrated for one million years using the
time-symmetrised Hermite algorithm\cite{kokubo1998} 
implemented
on graphics processing units (GPUs) in the Swarm-NG 
package\cite{swarm}
We found no
indications of orbital instability for any of the models considered and the
assumption of long-term orbital stability of the
three-body system does not provide additional constraint on the
current masses and orbital parameters of these systems.

For each of the three known circumbinary planets, we integrated an
ensemble of a few thousand three-body systems, each consistent with
the observed masses and orbital parameters, except that we varied the
semi-major axis of the planet.  We identify systems as unstable if the
planet's semi-major axis changes by more than 50\% from its original
value.  We report $a_{\rm min-stable}$, the minimum planetary semi-major axis
that was not flagged as unstable during the 10,000 year integrations.
The ratios of $a_{\rm min-stable}$ to the planets observed semi-major axes are
1.19 (Kepler-16b), 1.24 (Kepler-35) and 1.24 (Kepler-36).  The
corresponding ratios for the minimum stable planetary orbital period
to the planet's observed orbital period are 1.30, 1.38 and 1.37.

\section{Response of the planetary atmosphere to irradiation}

Circumbinary planets, as a class, will experience complex insolation
variations that may lead to climatic effects not expected in any other type
of planet. The radiative time constant over which an atmosphere radiates
away excess energy is approximately one month for the planets considered
here (see below), which would tend to smooth out the most rapid flux variation.
The advective timescale over which the atmosphere redistributes heat around
the planet is several days, indicating that the variable insolation should
lead to global, rather than local, changes in atmospheric temperature.
Transiting circumbinary planets will also likely experience frequent mutual
eclipses of their host stars causing a rapid decrease in the insolation for
a few hours; near 50\% decrease for Kepler-34. 

The radiative time constant of an atmosphere (the time to heat up or
cool off) can be estimated\cite{Showman02} to be 
\begin{equation}
\tau_{rad}
= {P \over g} {{c_p}\over {4 \sigma T^3}}, 
\end{equation}
where $P$
is the pressure, $g$ is the surface gravity, $c_p$ is the specific
heat capacity, $\sigma$ is the Stefan-Botzmann constant, and $T$ is
the temperature.  This equation is approximate, but is generally valid
at photospheric pressures.  Here we will choose $P= 1$ bar.  For these
Saturn-like exoplanets, the temperature at 1 bar should be near 500 K.
This yields $\tau_{rad}\sim 0.1$ years, or around one month.

The time scale for redistribution is the advective time scale,
$\tau_{adv}= R_{p} / U$, where $R_{p}$ is the planet radius, and $U$
is the wind speed.  Based on previous work modelling the dynamics of
giant exoplanet atmospheres, we expect a wind speed between 0.1-1
km s$^{-1}$ at 1 bar\cite{Showman10}.  Using $R_p= 7\times10^4$ km and
a wind speed of 0.3 km s$^{-1}$, this yields $\tau_{adv}\sim 3$ days.
The advective time is $\sim10 \times$ faster than the radiative time.
This shows efficient redistribution of absorbed
energy around the planet.

The finding that $\tau_{rad}$ is longer than a week, which is the
approximate period over which the incident flux varies dramatically,
means that this would tend to round out some of the severe climatic
disturbances driven by the incident flux changes.  However the short
$\tau_{adv}$ shows that the time-variable changes in climate that do
occur should be planet-wide in nature.

\section{The search for transiting circumbinary planet candidates}\label{search}

To determine what fraction of stars host Earth-like 
planets$^{11}$,
{\em 
Kepler} monitors the brightness of approximately 166,000 stars. As part of 
this exoplanet reconnaissance, 2165 eclipsing binaries are being observed 
of which 1322 are detached or semi-detached systems$^{13}$
We investigate 
these two subclasses of eclipsing binaries because the eclipse timing 
technique outlined in  Supplementary 
Section 8 does not work well if the first and 
fourth contact points (start of ingress and end of egress) are not well 
defined. We also chose to omit systems with $P<0.9$ days, as these in 
general also suffer from eclipse timing measurement difficulties
owing to out-of-eclipse variations due to tidal distortions and
reflection effects. Of the 
systems classified as detached or semi-detached with $P>0.9$ days, a total 
of 1039 systems have reliably measured orbital periods. 

For this investigation, out of the 1039 systems, we focus on 750
systems that exhibit primary and secondary eclipses. This requirement
for both eclipses to be present comes from the need to be able to
measure differences in orbital period defined by the primary eclipses
$P_{A}$ and the secondary eclipses $P_{B}$. We find this difference in
period to be the strongest indicator of a dynamical interaction with a
third body, especially in cases where the O--C variations are small.
The significance of the period difference accumulates in strength with
time while being insensitive to individual noise events.  Having both
primary and secondary eclipses is crucial, as otherwise one would
simply find no secular trend in the O--C diagram when only primary
eclipse times or secondary eclipse times are considered.  (It should
be noted that for circular orbits $P_{A} - P_{B} = 0$, so any
selection that relies purely on period differences will be biased
against finding third bodies if the EB stars are on circular orbits.)
The periods of these 750 systems range from 0.9 to 276 days, and these
data span a duration of 671 days.  Thus in the {\it Kepler} data there
are 750 systems with primary and secondary eclipses with $P$ ranging
from 0.9--276 days and classified as detached or semi-detached EBs.
This defines the sample used to search for transiting circumbinary
planets.


%

Of these 750 systems, 134 (18\%) exhibited greater than $3\sigma$ 
differences in primary and secondary orbital periods. Many of these showed 
large variations (tens of minutes to hours) and thus the perturbing body 
was presumed to be stellar in nature. The remaining systems with small 
timing variations could either have stellar-mass companions on distant 
orbits, or planet-mass companions in nearby orbits. Fortunately any 
periodicity in the O--C variations provides (usually within a factor of 2) 
the period of the 3rd body. The smallest variations with the shortest 
periods are therefore the most interesting when searching for circumbinary 
planets. However, this is also the regime where noise, and more seriously, 
spurious periodicities due to stellar pulsations and starspots, also 
affect the O-C curve, hampering the search.

Thus all 750 systems were examined for possible transit or tertiary 
eclipse events, not just the 134 most interesting cases. Since the 
presence of the primary and secondary eclipse precluded the use of 
standard planet-transit search algorithms, each light curve 
was inspected visually for the presence of transit events.
(Our initial attempt at fitting and removing the eclipses and then 
searching the residuals for transits did not work; there were always small 
remainders after the best-fit model was subtracted that would lead to 
spurious detections.) Planet transits-like events were found in four 
systems: KIC 8572936 (Kepler-34), KIC 9837578 (Kepler-35), KIC 12644769 
(Kepler-16), and KIC 5473556 (KOI-2939).

As described above, the search is neither fully complete nor fully 
quantifiable, and thus precludes a robust estimate on the frequency of 
circumbinary planets at the present time. However, a robust {\em lower 
limit} is possible, and is described in detail in the following section.

\section{The frequency of circumbinary planets}


There are several indications that the three observed transiting
circumbinary planets (TCBPs) are only a tiny fraction of circumbinary
planets, with the dominant reason being the geometric aspect: the
planets must be very well aligned to be seen in transit. Furthermore,
we have not searched all eclipsing binaries nor are we claiming that
these three planets are the results of an exhaustive search. In this
section, we estimate the geometric correction, but do not correct for
any search incompleteness or related factors, thus yielding a lower
limit circumbinary planet (CBP) frequency with approximately
order-of-magnitude level precision. Despite its limitations, the
estimated rate still provides significant insights into planet
formation around binary stars.

The combination of three-body interactions and radial velocity
measurements allow for a full measurement of the three-dimensional
orientation of the binary and planetary orbits. Using the known
orientation of the orbits (including the significant motion of the
stars around their barycentre) and an expansion of the technique in
ref.\ 21,
we can determine what fraction of randomly
placed observers would see these three systems eclipsing and
transiting, eclipsing and non-transiting, and non-eclipsing and
non-transiting. We describe three progressively more accurate ways of
estimating the geometric factors: the first technique treats the stellar
secondary as a planet, the second adds the barycentric motion of the
stars, and the third technique 
allows for non-coplanar orbits and is
calculated numerically.

The simplest model imaginable uses circular coplanar orbits where the
primary star is considered fixed as it is orbited by the secondary
star and the planet and we ignore planetary transits of the
secondary. In this approximation, the system is identical to the
multi-transiting systems discussed in
ref.\ 21.
The
probability that a binary undergoes eclipses is $(R_A+R_B)/a_1$ and the
probability that the planet transits given that the systems is
eclipsing is $a_1/a_2$ 
(ref.\ 21).
Therefore, the
geometric correction for the number of non-transiting planets where
the binary is eclipsing is 3.1, 4.8, and 3.4 times as many as
observed in the both transiting and eclipsing case for Kepler-16,
Kepler-34, and Kepler-35, respectively.

Improving this model requires accounting for the fact that the binary
stars sweep out a significant area as they move about their
barycentre (Figure 3), which we account for in this second technique.
The path on the sky of a circular orbit with semi-major axis $a$ is
an ellipse with major axis $a$ and minor axis $a \cos i$. Coplanar
orbits have zero mutual inclination ($\phi$). When $\phi$ is
non-zero, the
mutual inclination can be decomposed into contributions along the
line of sight (i.e., $i_2-i_1$) and in the plane of the sky
$\Omega_2$, as can be seen from the mutual inclination equation:
$\cos \phi = \cos i_1 \cos i_2 + \sin i_1 \sin i_2 \cos \Omega_2$. In
this technique, we assume fixed circular orbits with no mutual
inclination in the plane of the sky (difference of longitude of
ascending nodes $\Omega_2 = 0$) and no evolution of the two-body
orbital elements given in Table 1.

In this case, the orbital paths of the three bodies form concentric
ellipses with major axes corresponding to the semi-major axis
measured with respect to the barycentre, which we will approximate as
$a'_A = a_1 (M_B/(M_A+M_B))$, $a'_B = a_1 (M_A/(M_A+M_B))$ and
$a'_p=a_2$.  In this case, transits of the planet over the primary
can occur when the semi-minor axis of the planet's orbit is less than
the semi-minor axis of the primary's apparent orbit plus the sum of
the radii of the bodies, i.e., $a'_p \cos i_2 < a'_A  \cos i_1 + R_A
+ R_p$. In coplanar systems ($i = i_1 = i_2$), this criterion becomes
$\cos i < (R_A+R_p)/(a'_p - a'_A)$; since random orientations imply a
uniform distribution in $\cos i$, the probability of transit is
$(R_A+R_p)/(a'_p-a'_A)$. This is to be compared to the probability of
transit if the primary was fixed, which would be $(R_A+R_p)/a'_p$.
Given that $a'_p$ is often rather larger than $a'_A$, we can Taylor
expand this expression to get an enhancement factor of approximately
$1 + (a'_A/a'_p)$ in the probability of orbit crossing due to the
fact that the secondary is moving the primary around its barycentre
(again in the circular coplanar case). Similar expressions can be
derived for crossings of the planet across the secondary's orbit. The
secondary has a larger orbit ($a'_B > a'_A$) but usually a smaller
radius, so orbit crossings of the secondary should also be evaluated.
For these three systems, under these approximations, the motion of
the binary around its barycentre increases the probability of orbit
crossing by about 25\%. If the planet does not have a resonant
relationship with the binary, then eventually all objects will
explore all phases and on long timescales, and these orbit crossing
criteria can be called transit criteria. (On long-time scales, the
inclinations can change, but the appropriate way of determining
the frequency of circumbinary planets is to fix the observed
inclination to the inclination at the time of discovery.)

When there is a mutual inclination that is not entirely towards the
line of sight, then the orbital tracks of the three objects remain
the same, except with a rotation between the planetary ellipse and
the binary ellipse by the difference in the longitude of ascending
nodes ($\Omega_2$), as can be seen in Figure 3. For low values of
$\Omega_2$ the above approximations are still mostly valid. However,
calculating the exact close approach distance between two non-aligned
concentric ellipses is more accurate; this is most easily calculated
numerically. 
A Monte Carlo code$^{21}$
was developed that uses the full three dimensional orientation of the
orbits (retaining the assumption of circular fixed orbits of each
object around the barycentre) and places random observers
isotropically on the sphere. Each observer either sees the system
non-eclipsing and non-transiting, eclipsing and non-transiting, or
eclipsing and transiting, where we call a system ``transiting'' if the
on-the-sky projection of the planetary orbit and the orbit of the
primary or the secondary have a close approach distance less than the
sum of the radii, though this does not guarantee a transit every time
this close approach distance is reached by the planet. These
approximations are sufficient for the order-of-magnitude lower-limit
rate estimates we are considering here.

Applying this model to Kepler-16, Kepler-34, and Kepler-35 to correct
for geometric completeness results in approximately 5, 9, and 7 times
as many EBs which have CBPs (most non-transiting) and approximately
260, 180, and 150 times as many binaries that are non-eclipsing and
non-transiting, respectively. (There is a small difference
($\lesssim$20\%) in these numbers with or without including crossings
of the secondaries for these three systems which we ignore.) So, if
all the EBs were analogues to the three observed systems, we would
expect at least $\sim21$ ($5+9+7$) 
{\em Kepler} EBs have CBPs, most of which would
be non-transiting. 


In reality, the EB sample is not similar to analogues of these three EBs
as most EBs have shorter periods than those we see here; only 133 of the
750 searched systems have periods greater than 20 days. To zerorth order,
the probability of detecting a coplanar transiting CBP at a fixed period
(e.g., the 100-200 day periods for these systems) is equally likely for
an EB of any period. (Though there are many more EBs at shorter periods,
most of these are not aligned to within 0.5 degrees that is required for
transiting systems.) That all three detections came from the small sample
of longer-period binaries is very suggestive that 100-200 day period
planets are not equally present around binaries of all periods. Drawing a
firmer conclusion will only be possible with additional work, since, to
first order, tighter binaries have smaller transit enhancements from
barycentric motions and also spend much more time in eclipse, when
transits are much more difficult to detect. Thus, keeping in mind that an
exhaustive search has not been completed, it is possible that the
discovery of three planets at periods greater than 20 days is because
these systems are slightly more likely to reveal CBPs, along with small
number statistics.


If we consider CBPs at scaled periods near the dynamical stability limit,
like the ones observed here, then the likelihood of finding transiting
CBPs around short-period binaries is much higher since these planets
would have shorter periods and many more transits. It is therefore
interesting, but again not conclusive, that the first CBPs were not found
around shorter-period binaries, suggesting that shorter-period binaries
have a much lower rate of gas giant CBPs near the dynamical stability
limit. This will be clarified in future work. Note that restricting the
calculation to binaries with periods between 20 and 50 days would cause
the rate to go up by a factor of $750/133 \sim 6$ (though the total
number of such CBPs in the galaxy will only increase by at most a factor
of 2 since there are three times fewer binaries in this 20-50 day period
range, see below.) However, we will not restrict the period range in our
calculation of the CBP frequency, preferring to use the entire sample of
the 750 searched EBs.


Returning to the case of planets with periods like the ones observed, the
detection of the known planets around the 750 searched EBs would yield a
smaller rate than observing the known planets around 750 analogues of the
current systems (because transits are slightly less frequent and harder
to detect as the EB period decreases). Thus, we can use the latter
distribution to determine an underestimate of the frequency of CBPs of
approximately $21/750 = 3\%$. (The former distribution would have a
smaller denominator when considering that the shorter period systems
would have lower detection probabilities and thus lower weight).

Small number statistics suggest that if the probability of a planet
transiting when eclipsing is $\sim1/6$, then observing 3 systems is
consistent with the true rate being $21^{+20}_{-12}$, so that the
one-sigma lower limit rate is $9/750 = 1.2\%$. As discussed above, this
lower limit is an underestimate since the three known systems are not
the final result of an exhaustive search.

The geometrical arguments presented above can also be considered by
looking at a fraction of \emph{Kepler} stars instead of just
\emph{Kepler} EBs, though these are not independent arguments. We find
that approximately $260+180+150=590$ systems in the \emph{Kepler} field are
non-transiting and non-eclipsing binaries with CBPs. Using a binary
fraction of sunlike stars of 44\% and that 6\% of these have orbital
periods between 0.9 and 50 days (see below; 
ref.\ 2),
suggests that 
$160000 \times 0.44 \times 0.06 = 4200$ 
\emph{Kepler} targets are qualifying binaries, resulting in
a frequency estimate of roughly $590/4200 \sim 10\%$. However, this
calculation does not account for the period distribution of binaries or
for the differences between the binary fraction of \emph{Kepler} targets
and the volume-limited survey of ref.\ 2, 
which would lower the estimated
frequency.


Using these geometric arguments, we claim a lower-limit frequency of
circumbinary planets like those presented here (i.e., Saturn-like,
periods around 100-200 days) of 
$\sim 1\%$ of binaries with periods between 0.9
and 50 to order of magnitude precision. This is similar to the rate
of planets on 100-365 day periods around single stars from radial
velocity surveys\cite{cumming2008}
and the frequency of planets around
members of wide binaries is also known to be similar$^2$.
The properties of this new class of circumbinary gas giant planets
will be a challenge for planet formation theories; Kepler-34 and
Kepler-35 show that such planets are relatively common and can exist
around binaries with a variety of eccentricities, masses, mass
ratios, and average insolation.

The duration of {\em Kepler} 
observations investigated for this study is
670.8 days, so to guarantee two transits, period of planet must be
less than $670.8/2=335.4$ days. (Alternatively, it is straightforward
to show using a one-dimensional geometric argument that the period
for which a randomly chosen epoch will have two transits 50\% of the
time is just the duration of the observations.) To ensure long-term
dynamical stability$^{17}$,
a period ratio between the binary and the planet
should be greater than about 5 (5.6 is the smallest seen here). 
Using the observed period ratio range of 5-10,
we can say that binaries with periods of less than 34 days would have
clearly had two passes by a planet near the dynamical stability
limit, with some residual sensitivity up to binary periods of 134
days. For the purposes of discussing CBP rates, we will combine the
original search criterion of $P>0.9$ days (to focus on detached
binaries) with an upper limit of about 50 days, where our sensitivity
starts to drop.

Most of the stars in the Milky Way are in the Milky Way disk, whose
mass is not well known.  One of the lower estimates suggests
that the Milky Way disk contains roughly $10^{10.5}$
solar masses, about half of which is in stars and half
in the interstellar medium\cite{binney}.
Most of the mass in the stellar component is in 
sun-like stars, implying there are approximately
$10^{10}$ sun-like stars in the Galaxy$^{23}$
%
A recent
solar-neighbourhood volume-limited survey$^2$
found that 44\% of FGK stars are binaries, with a log-normal
distribution in period (mean of 
$\log P = 5.03$ and standard deviation
of $\sigma_{\log P} = 2.28$), with period in days, suggesting that 5.9\% of
binaries have periods between 0.9 and 50 days. Thus, the number of
sunlike stars that are binaries with periods between 0.9 and 50 days
in the Milky Way is roughly 
$10^{10} \times 0.44 \times 0.059 \approx 10^{8.5}$. 
Assuming no significant difference between {\em Kepler}
stars and stars in the Galaxy, our lower-limit circumbinary planet frequency
estimate of 1\% suggests that there are several million circumbinary
planets like the ones we discovered here in the Milky Way.


\section{Supplementary notes}

\spacing{1}

\newpage

\section{Supplementary figures, legends, and tables}

~

\newpage

\newpage

\newpage

\begin{figure}
\centerline{\includegraphics[scale=0.7,angle=-90]{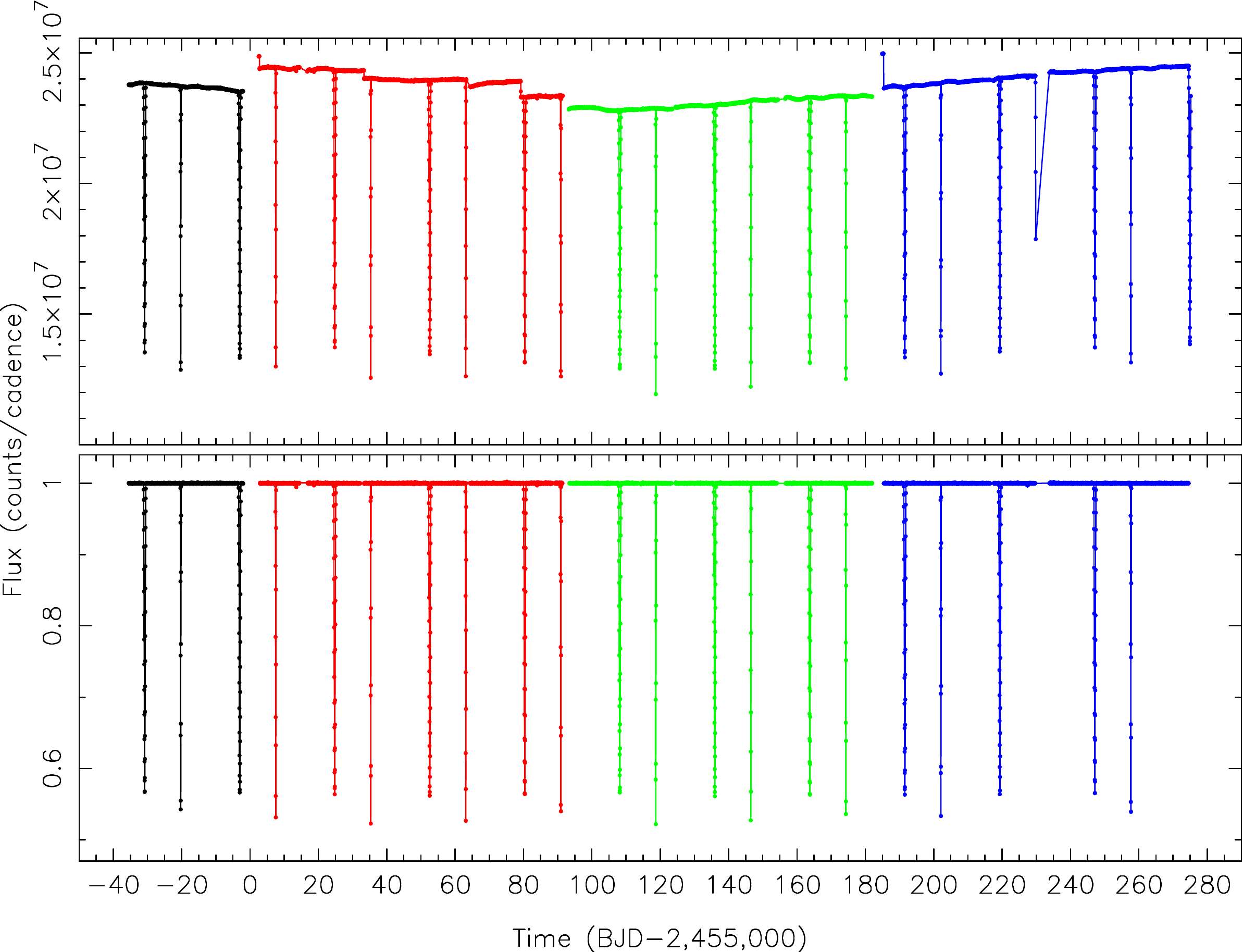}}
\caption{{\bf $\vert$ Light curve detrending for 
Kepler-34.}
Top: The ``PA'' light curves for Kepler 34 are shown for quarters
Q1 (black) through Q4 (blue).  
The Q2 light curve (in red) shows
some instrumental artefacts in the out-of-eclipse regions, including
short-term sensitivity changes and drifts due to spacecraft
pointing adjustments.
A primary eclipse was interrupted by a data gap in the middle
of Q4, and a secondary eclipse was interrupted by the ending of Q4.  
Apart
from the instrumental artefacts, there is little out-of-eclipse
variability on this scale.
Bottom:  The detrended and normalized light curve.  The partially
observed primary and secondary eclipses
in Q4 were removed.
The light curves
from other quarters were also detrended, but are not shown here
for the sake of clarity.
\label{plotraw2459}}
\end{figure}

\newpage
\begin{figure}
\centerline{\includegraphics[scale=0.7,angle=-90]{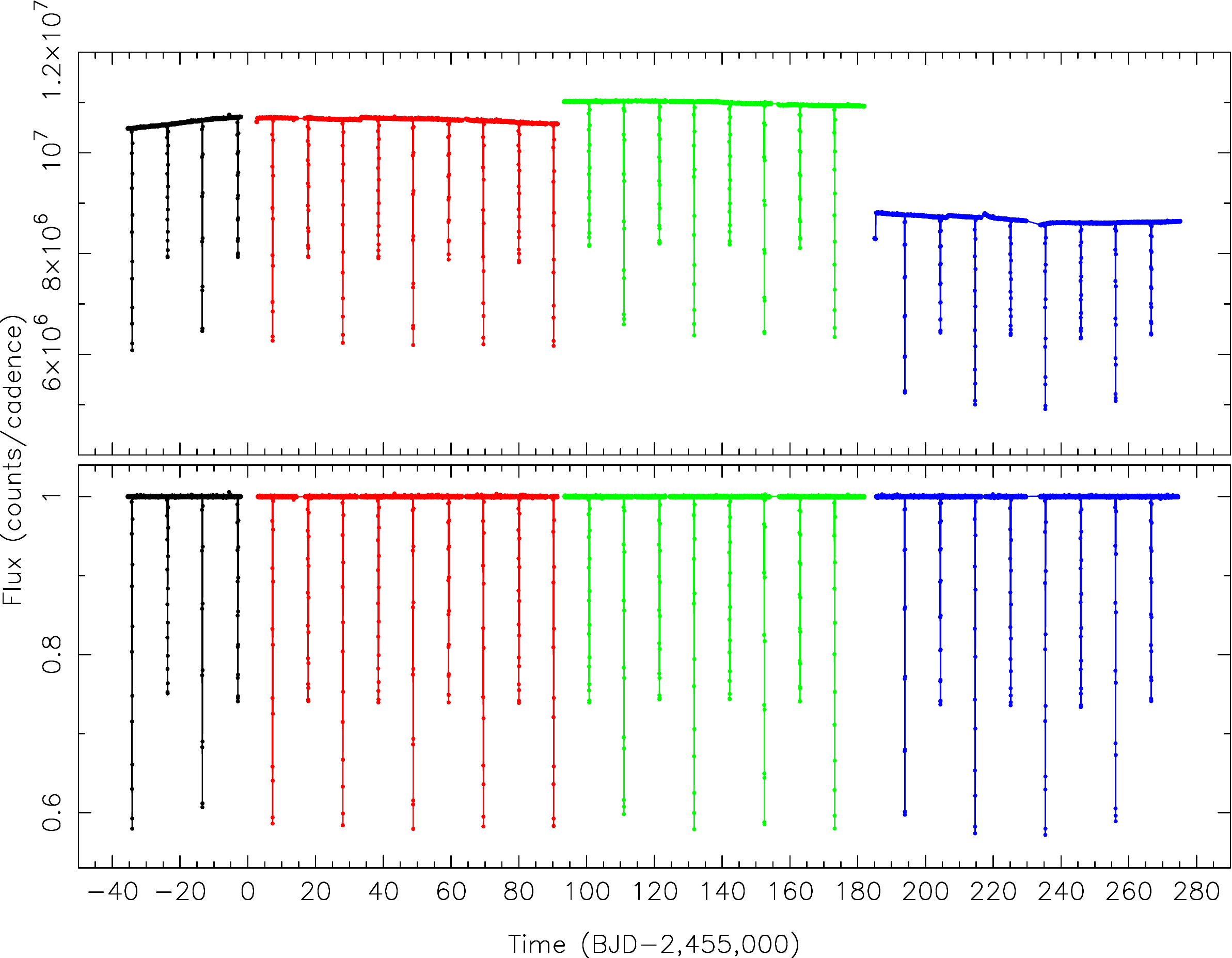}}
\caption{{\bf $\vert$ Light curve detrending for 
Kepler-35.}
Top: The ``PA'' light curves for Kepler 35 are shown for quarters
Q1 (black) through Q4 (blue).   The instrumental artefacts
here are not as large as they are for Kepler 34
(Supplementary
Figure S\ref{plotraw2459}).  Apart
from the instrumental artefacts, there is little out-of-eclipse
variability on this scale.  One secondary eclipse was missed in the gap between
Q3  (green) and Q4.
Bottom:  The detrended and normalized light curve.  
The light curves
from other quarters were also detrended, but are not shown here
for the sake of clarity.
\label{plotraw2937}}
\end{figure}

\newpage

\begin{figure}
\centerline{\includegraphics[scale=0.62]{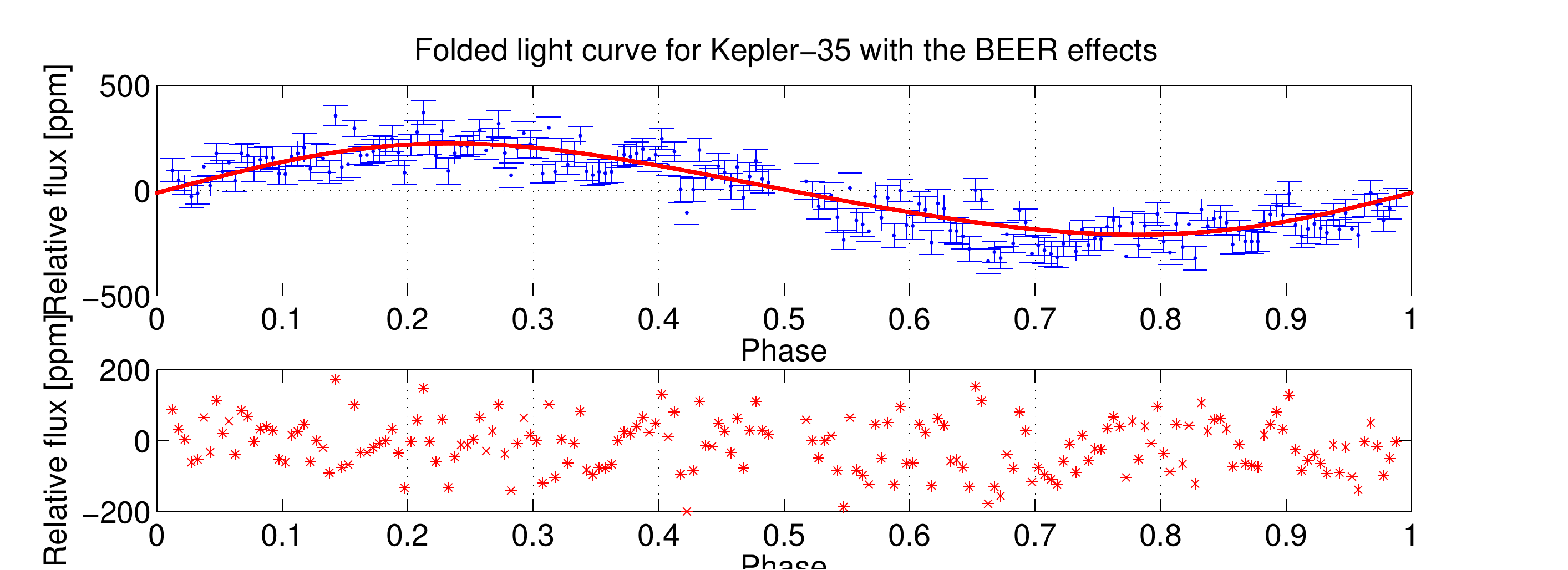}}
\caption{{\bf $\vert$ 
Doppler beaming effect in Kepler-35.}
Folded, cleaned, out-of-eclipse 
light curves, binned into 200 bins, of Kepler-35.
Phase zero is mid primary eclipse
in this figure. The errors of each bin represent
$1\sigma$ estimate the  bin average value. The line presents the
Doppler 
beaming  model. The model residuals are plotted in the lower panel.
\label{model_fit}}
\end{figure}

\newpage

\begin{figure}
\centerline{\includegraphics[scale=0.7,angle=-90]{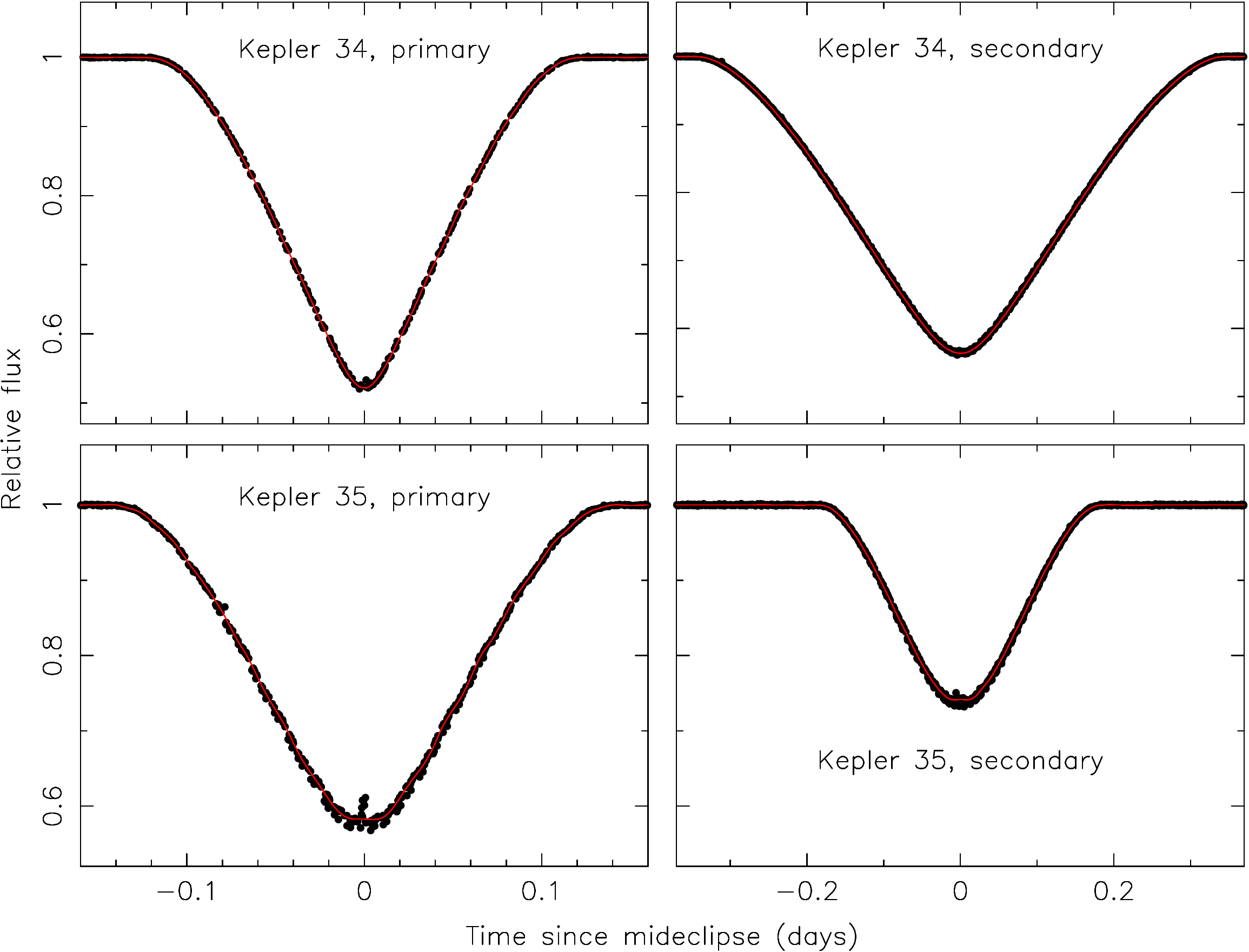}}
\caption{{\bf $\vert$ Eclipse profiles for 
Kepler-34 and Kepler-35.}
The folded primary and secondary eclipses for Kepler-34 and
Kepler-35 (filled circles) with the template profiles used
to measure times of mideclipse for each event (solid lines).
The few bright points near the middle of primary eclipse
in both sources are artefacts caused by the cosmic ray
rejection software in the {\em Kepler} data analysis pipeline.
\label{plotprof}}
\end{figure}

\newpage

\begin{figure}
\centerline{\includegraphics[scale=0.7,angle=-90]{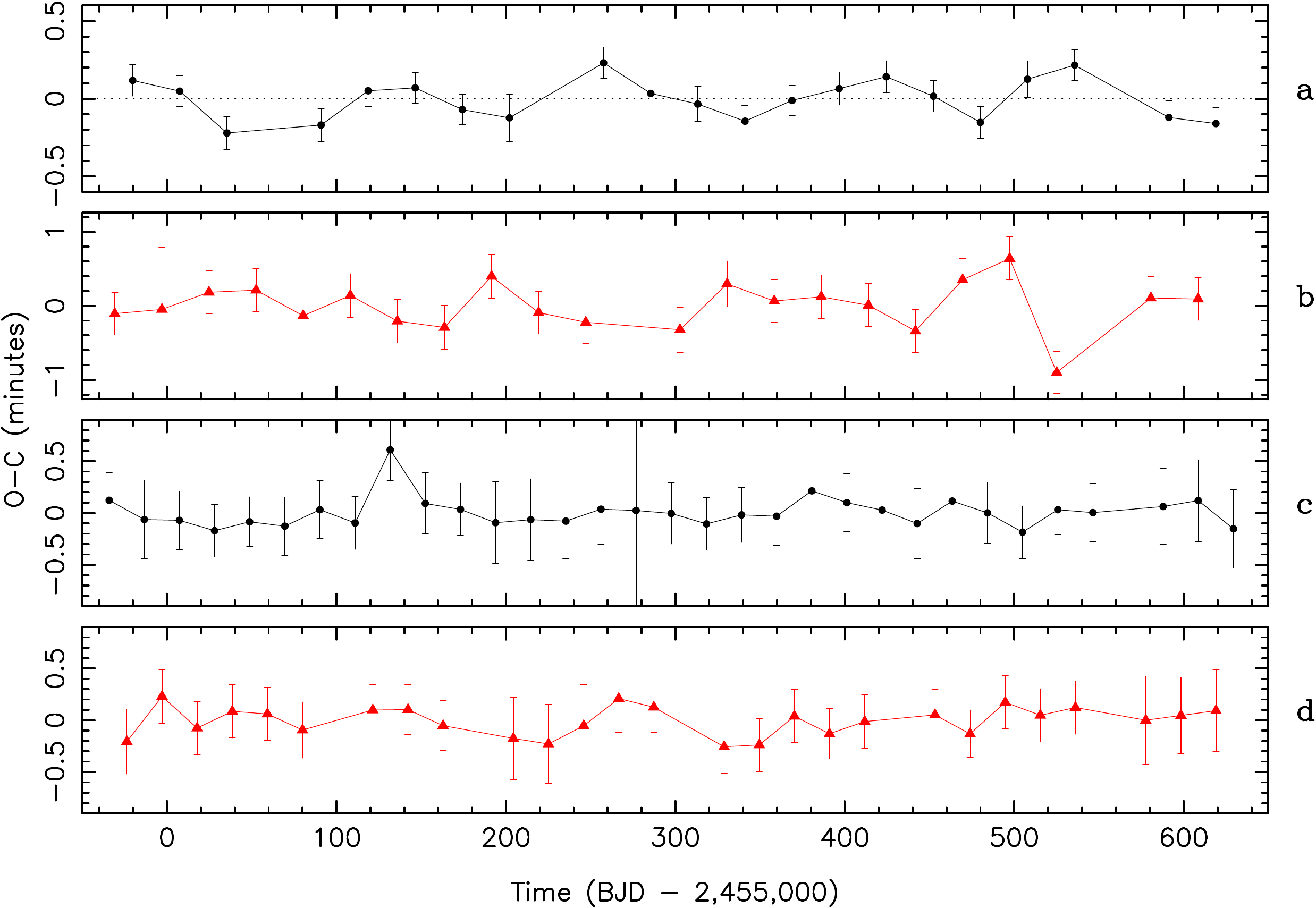}}
\caption{{\bf $\vert$ O--C diagrams for
Kepler-34 and Kepler-35.}
Observed-Computed (O--C) diagrams for the Kepler-34 primary
eclipse times (a), secondary eclipse times (b), Kepler-35
primary eclipse times (c), and secondary eclipse times (d).
\label{plotoc}}
\end{figure}

\newpage

\begin{figure}
\centerline{\includegraphics[scale=0.66,angle=0]{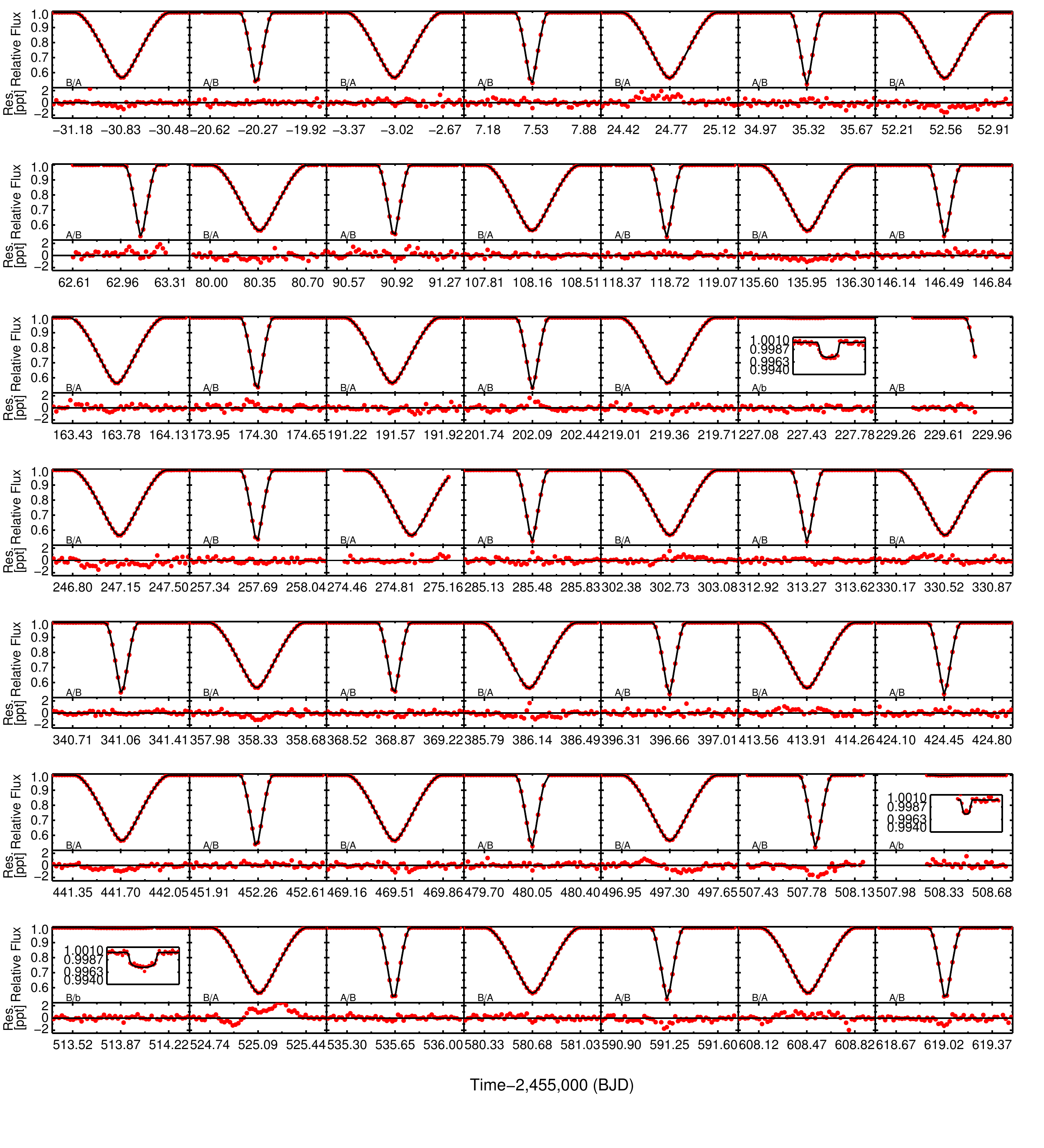}}
\vspace{-2em}
\caption{{\bf $\vert$ Light curves and photodynamical model for 
Kepler-34.}
Individual eclipse events for Kepler-34 (red circles) and the
best-fitting photodynamical model (black line).  
Primary
eclipses are marked with ``A/B'' and secondary eclipses
marked with ``B/A''.  Planet crossings of the primary star
are marked with ``A/b'' and planet crossings of the
secondary star are marked with ``B/b''.
The corresponding
residuals are shown in the thin panels below each eclipse plot.
The large residuals seen in the primary eclipse near day 525.09
are most likely due to a spot crossing the primary during the eclipse.
\label{output857}}
\end{figure}

\newpage

\begin{figure}
\centerline{\includegraphics[scale=0.66,angle=0]{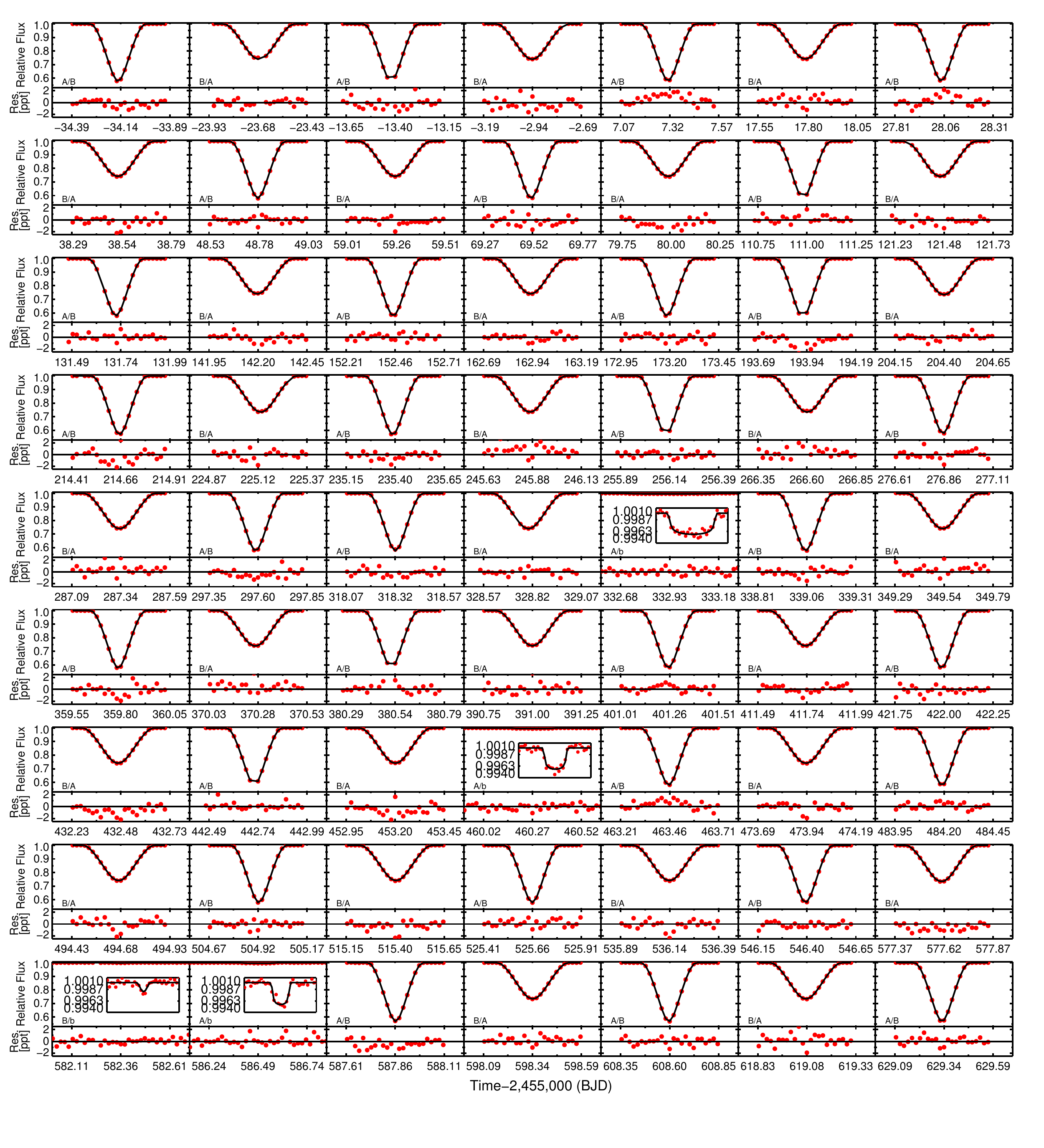}}
\vspace{-2em}
\caption{{\bf $\vert$ Light curves and photodynamical model for 
Kepler-35.}
Individual eclipse events for Kepler-34 (red circles) and the
best-fitting photodynamical model (black line).  
Primary
eclipses are marked with ``A/B'' and secondary eclipses
marked with ``B/A''.  Planet crossings of the primary star
are marked with ``A/b'' and planet crossings of the
secondary star are marked with ``B/b''.
The corresponding
residuals are shown in the thin panels below each eclipse plot.
\label{output983}}
\end{figure}

\newpage

\begin{figure}
\centerline{\includegraphics[scale=0.65]{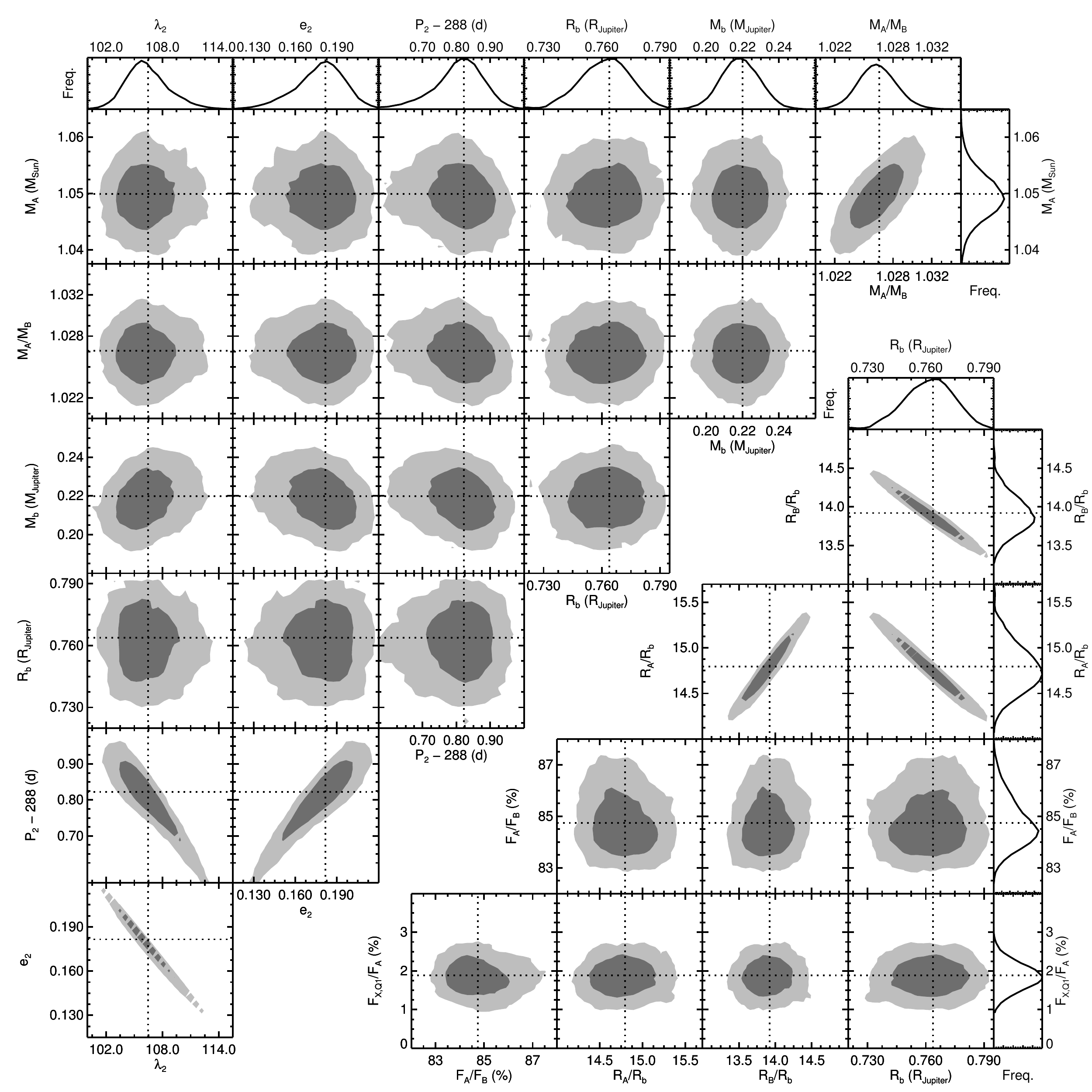}}
\caption{{\bf $\vert$ MCMC parameter correlations
for Kepler-34.}
Two-parameter joint posterior distributions for a selection of model
parameters. The
68\% and 95\% confidence regions are denoted by dark and light gray
shaded areas, respectively.
Single parameter marginalised distributions are plotted at the top
and/or to the far right of the
panels. The dashed lines mark the median values of the marginalised
distributions of each
parameter.
\label{mcmc857}}
\end{figure}

\newpage

\begin{figure}
\centerline{\includegraphics[scale=0.65]{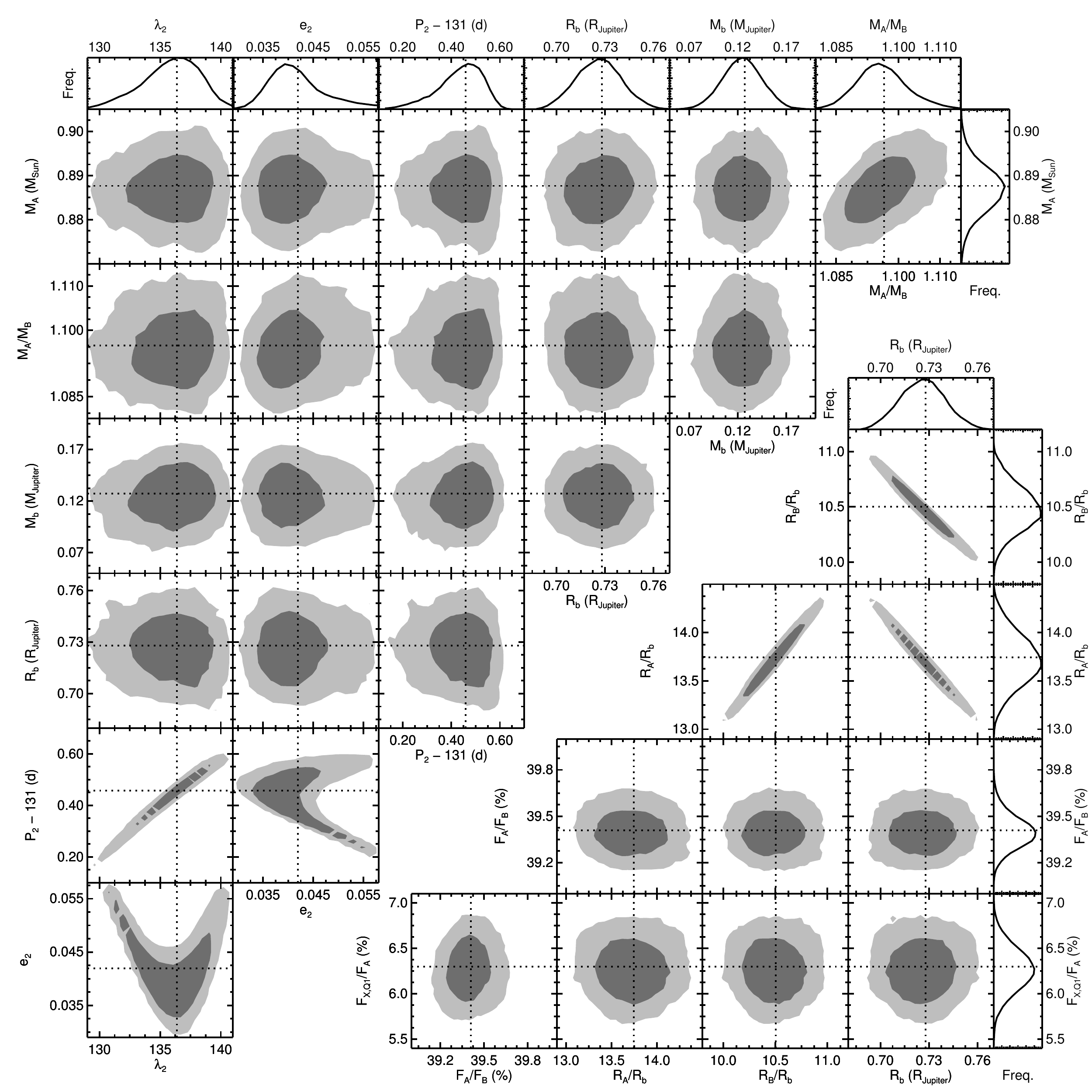}}
\caption{{\bf $\vert$ MCMC parameter correlations
for Kepler-35.}
Similar to Supplementary Figure S\protect\ref{mcmc857}, but
for Kepler-35.
\label{mcmc983}}
\end{figure}

\newpage

\begin{figure}
\centerline{\includegraphics[scale=0.9]{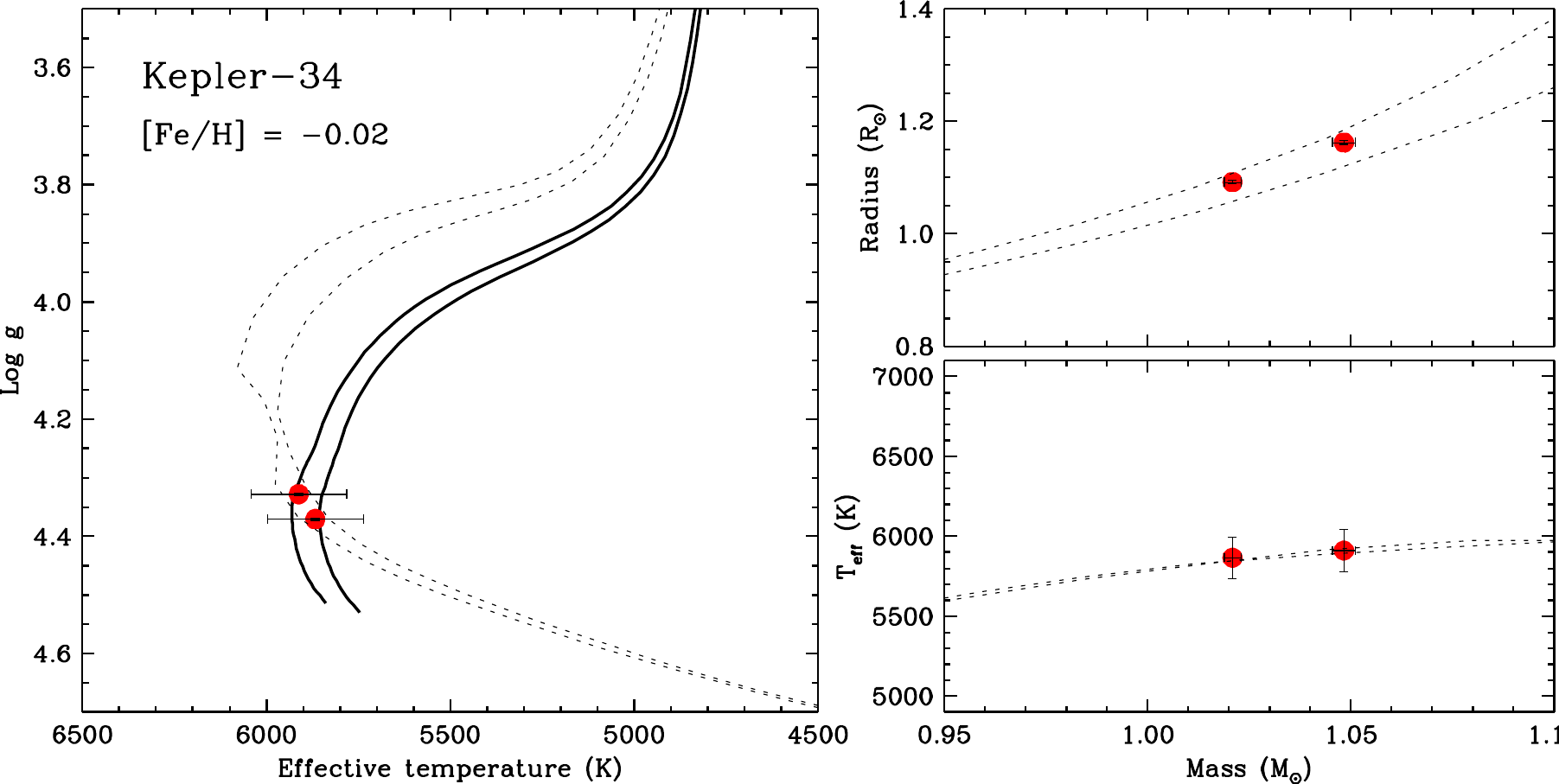}}
\caption{{\bf $\vert$ Isochrones for Kepler-34.}
Left: A $\log g$ versus effective temperature diagram showing
the measurements for Kepler-34. Evolutionary tracks$^{16}$
for
the measured masses are depicted with solid lines, for a metallicity
of 
${\rm [Fe/H]} = -0.02$ that provides the best fit to the measured
temperatures. The dotted lines represent isochrones for ages of 5 Gyr
(lower) and 6 Gyr, and the same metallicity.  Right: Mass-radius and
mass-temperature diagrams showing the measurements and the same two
isochrones as in the left panel.
\label{KOI2459iso}}
\end{figure}

\newpage

\begin{figure}
\centerline{\includegraphics[scale=0.9]{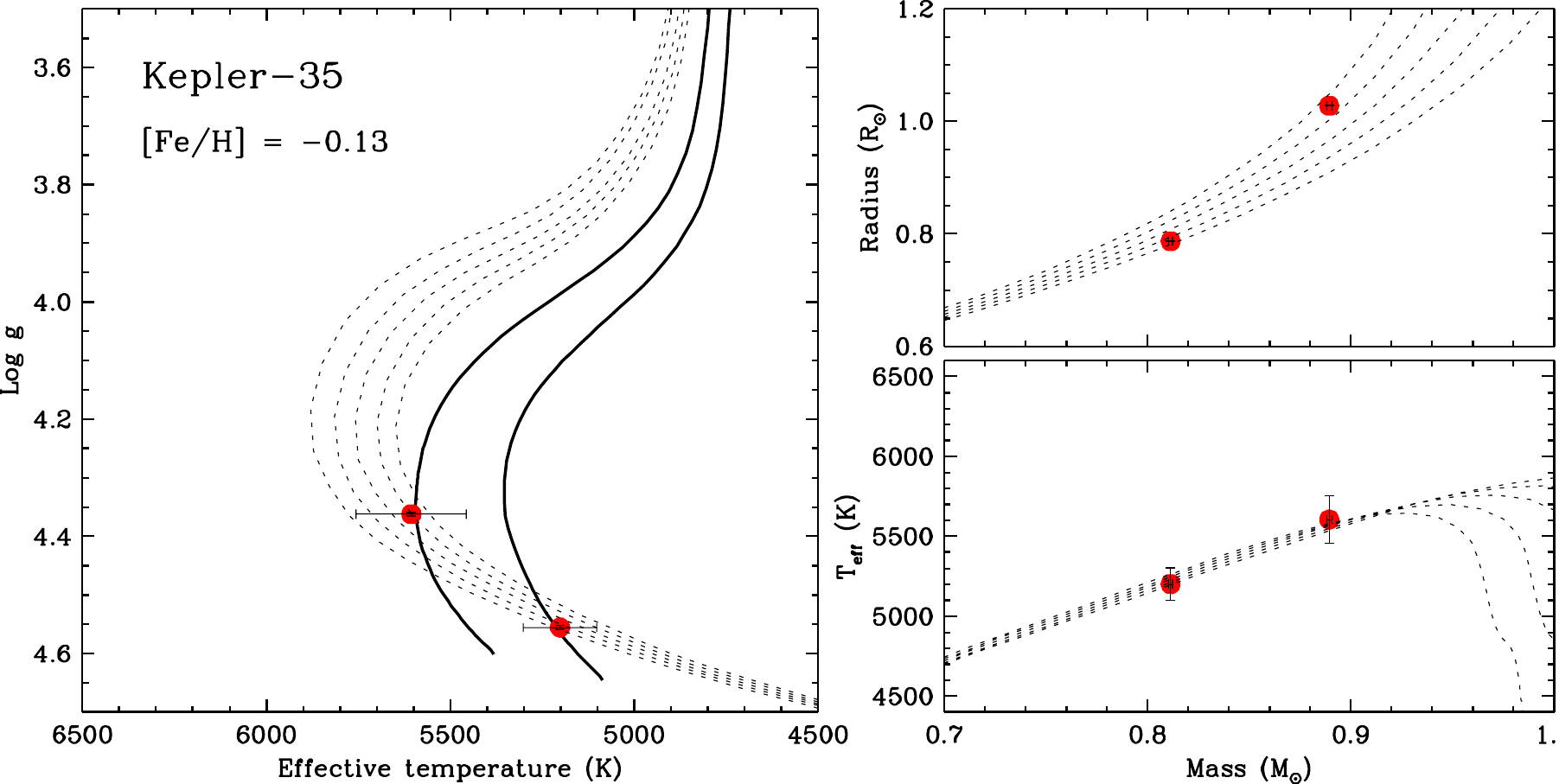}}
\caption{{\bf $\vert$ Isochrones for Kepler-35.}
Same as Supplementary Figure \protect S\ref{KOI2459iso}, for
Kepler-35.
\label{KOI2937iso}}
\end{figure}

\newpage

\begin{figure}
\centerline{\includegraphics[scale=0.85]{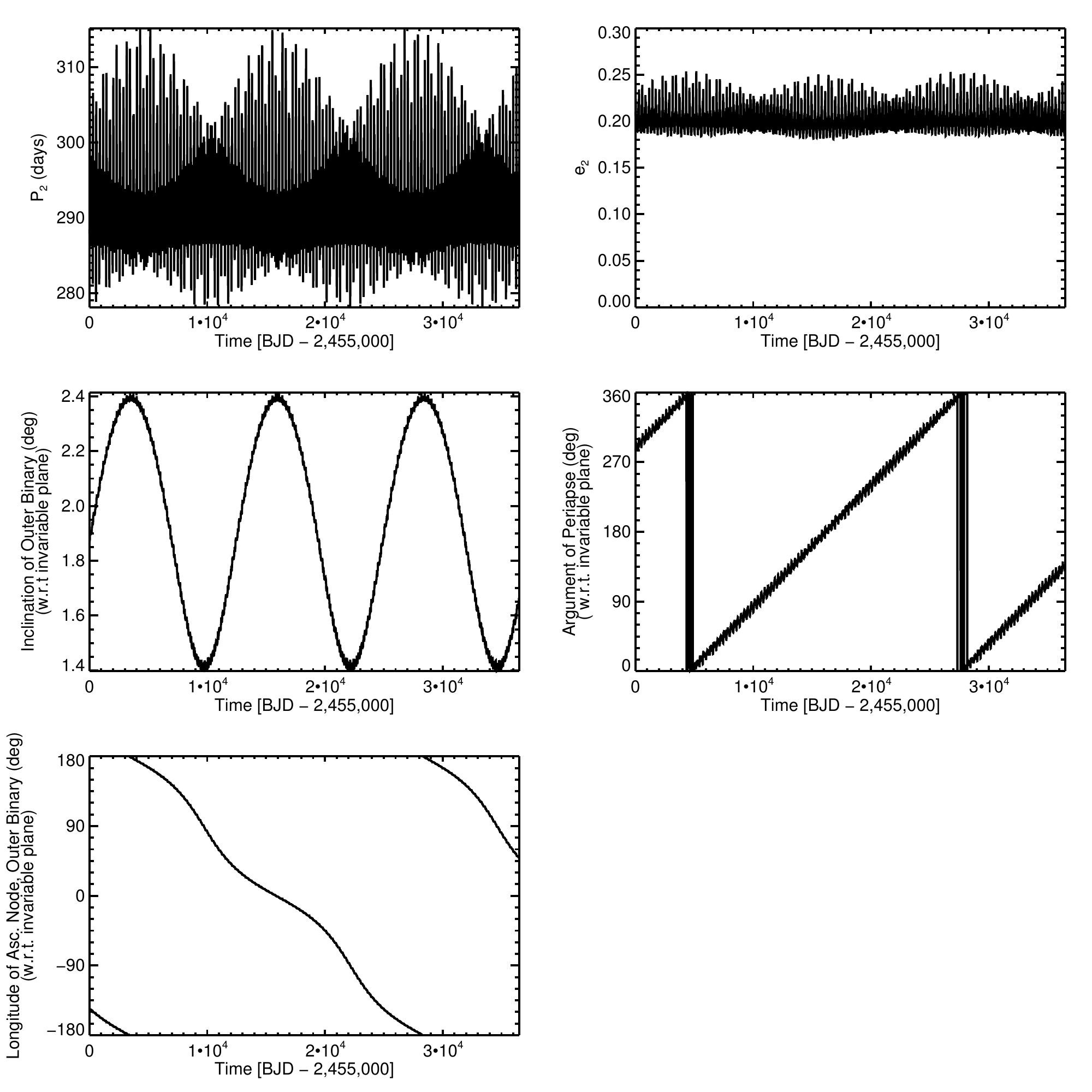}}
\caption{{\bf $\vert$ Evolution of the orbital elements
for Kepler-34.}
The evolution of the period of Kepler-34b, its eccentricity, inclination
relative to the stellar binary orbital plane, argument of periastron, and
its longitude of ascending node over a 100 year baseline.
\label{els857}}
\end{figure}

\newpage

\begin{figure}
\centerline{\includegraphics[scale=0.85]{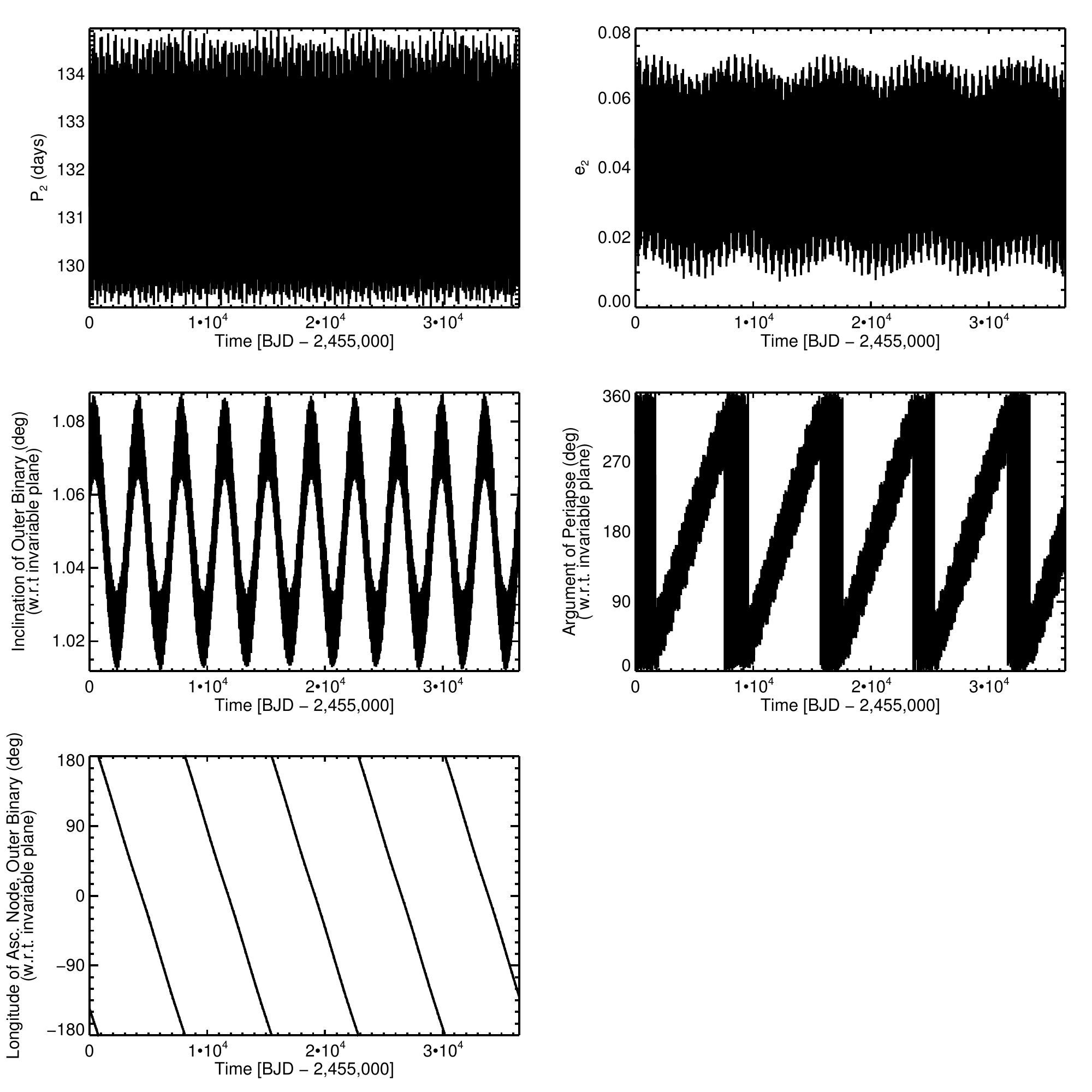}}
\caption{{\bf $\vert$ Evolution of the orbital elements
for Kepler-35.}
The evolution of the period of Kepler-35b, its eccentricity, inclination
relative to the stellar binary orbital plane, argument of periastron, and
its longitude of ascending node over a 100 year baseline.
\label{els983}}
\end{figure}

\newpage

\begin{table}
\centering
{\small
\begin{tabular}{lll}
\hline  & {\bf Kepler-34} & {\bf Kepler-35}\\ \hline 
\multicolumn{3}{c}{{\bf System Properties}} \\ \hline
KIC designation & 8572936   &  9837578  \\
KOI number      & 2459      &  2937     \\
2MASS designation  & 19454459+4438296 &      19375927+4641231 \\
right ascension (HH:MM:SS.S)  & 19:45:44.6 & 19:37:59.3   \\
declination     (DD:MM:SS.S)  & +44:38:29.6  & +46:41:23.6 \\
equinox              & 2000.0   &  2000.0   \\
Kepmag         & 14.875 &   15.726  \\
$J$ magnitude         & 13.605 &   14.425  \\
$E(B-V)$ (mag)        & 0.148  &  0.123     \\
\hline\hline
\multicolumn{3}{c}{{\bf Planetary Properties}} \\ \hline
Mass of planet, $M_p (M_{\rm Jupiter})$ & $ 0.220_{-0.010}^{+0.011} $ &  $ 0.127_{-0.020}^{+0.020} $ \\ 
Radius of planet, $R_p (R_{\rm Jupiter})$ & $ 0.764_{-0.014}^{+0.012} $ &  $ 0.728_{-0.014}^{+0.014} $ \\ 
Mean density of planet, $\rho_p$ (g cm$^{-3}$) & $ 0.613_{-0.041}^{+0.045} $ &  $ 0.410_{-0.069}^{+0.070} $ \\ 
Planet surface gravity, $g_b$ (cm s$^{-2}$) & $936_{-54}^{+57}$ & $596_{-98}^{+98} $ \\
\hline \hline
\multicolumn{3}{c}{{\bf Properties of the Planetary Orbit}} \\ \hline
Reference epoch (BJD)& 2,454,969.20000 & 2,454,965.85000 \\
Period, $P$ (days) & $ 288.822_{-  0.081}^{+  0.063} $ &  $ 131.458_{-  0.105}^{+  0.077} $ \\ 
Semi-major axis length, $a$ (AU) & $ 1.0896_{-0.0009}^{+0.0009} $ &  $ 0.60347_{-0.00103}^{+0.00101} $ \\ 
Eccentricity, $e$ & $ 0.182_{-0.020}^{+0.016} $ &  $ 0.042_{-0.004}^{+0.007} $ \\ 
Eccentricity times sine of arg.\ of periapse, $e \sin(\omega)$ & $ 0.025_{-0.007}^{+0.007} $ &  $ 0.035_{-0.011}^{+0.009} $ \\ 
Eccentricity times cosine of arg.\ of periapse, $e \cos(\omega)$ & $ 0.180_{-0.021}^{+0.016} $ &  $ 0.017_{-0.018}^{+0.021} $ \\ 
Mean longitude, $\lambda \equiv M+\omega$ (deg) & $ 106.5_{-  2.0}^{+  2.5} $ &  $ 136.4_{-  2.7}^{+  2.1} $ \\ 
Inclination $i$ (deg) & $ 90.355_{- 0.018}^{+ 0.026} $ &  $ 90.76_{- 0.09}^{+ 0.12} $ \\ 
Relative nodal longitude, $\Omega$ (deg) & $ -1.74_{- 0.16}^{+ 0.14} $ &  $ -1.24_{- 0.33}^{+ 0.24} $ \\ \hline \hline
\multicolumn{3}{c}{{\bf Properties of the Stellar Binary Orbit}} \\ \hline 
Reference epoch (BJD)& 2,454,969.20000 & 2,454,965.85000 \\
Period, $P$ (days) & $   27.7958103_{-   0.0000015}^{+   0.0000016} $ &  $   20.733666_{-   0.000012}^{+   0.000012} $ \\ 
Semi-major axis length, $a$ (AU) & $ 0.22882_{-0.00018}^{+0.00019} $ &  $ 0.17617_{-0.00030}^{+0.00029} $ \\ 
Eccentricity, $e$ & $ 0.52087_{-0.00055}^{+0.00052} $ &  $ 0.1421_{-0.0015}^{+0.0014} $ \\ 
Eccentricity times sine of arg.\ of periapse, $e \sin(\omega)$ & $ 0.49377_{-0.00060}^{+0.00057} $ &  $ 0.1418_{-0.0015}^{+0.0014} $ \\ 
Eccentricity times cosine of arg.\ of periapse, $e \cos(\omega)$ & $ 0.165828_{-0.000061}^{+0.000065} $ &  $ 0.0086413_{-0.0000031}^{+0.0000031} $ \\ 
Mean longitude, $\lambda \equiv M+\omega$ (deg) & $ 300.1970_{-  0.0105}^{+  0.0099} $ &  $  89.1784_{-  0.0012}^{+  0.0011} $ \\ 
Inclination $i$ (deg) & $  89.8584_{-  0.0083}^{+  0.0075} $ &  $  90.4238_{-  0.0073}^{+  0.0076} $ \\ 
Mean primary eclipse period (days)& $27.7958070\pm 0.0000023$  & $20.7337496 \pm 0.0000039$ \\
Mean secondary eclipse period (days) & $27.7957502\pm 0.0000065$  & $20.7337277 \pm 0.0000040$  \\
Reference time for primary eclipse (BJD-2,400,000) & $54979.72308\pm0.000036$ & $54965.84580\pm0.000034$\\
Reference time for secondary eclipse (BJD-2,400,000) & $54969.17926\pm0.000085$& $54976.32812\pm0.000033$\\
\hline 
 \end{tabular}
	}
\caption{$\vert$ A summary of system information for Kepler-34 and
Kepler-35 taken from the KIC, and a summary 
of the planetary
properties, the planetary orbit, and the stellar binary orbit
determined by the photometric-dynamical model and eclipse timing
analysis.\label{tab1}}
\end{table}

\newpage

\begin{table}
\centering
{\small
\begin{tabular}{lll}
\hline  & {\bf Kepler-34} & {\bf Kepler-35}\\ \hline 
\multicolumn{3}{c}{{\bf Properties of the Stars in the Stellar Binary}} \\ \hline 
Mass of primary, $M_A (M_{\odot})$ & $ 1.0479_{-0.0030}^{+0.0033} $ &  $ 0.8877_{-0.0053}^{+0.0051} $ \\ 
Radius of primary, $R_A (R_{\odot})$ & $ 1.1618_{-0.0031}^{+0.0027} $ &  $ 1.0284_{-0.0019}^{+0.0020} $ \\ 
Mass of secondary, $M_B (M_{\odot})$ & $ 1.0208_{-0.0022}^{+0.0022} $ &  $ 0.8094_{-0.0045}^{+0.0042} $ \\ 
Radius of secondary, $R_B (R_{\odot})$ & $ 1.0927_{-0.0027}^{+0.0032} $ &  $ 0.7861_{-0.0022}^{+0.0020} $ \\ 
Primary surface Gravity, $\log g_A$ [cgs] & $ 4.3284_{-0.0019}^{+0.0023} $ &  $ 4.3623_{-0.0020}^{+0.0020} $ \\ 
Secondary surface Gravity, $\log g_B$ [cgs] & $ 4.3703_{-0.0024}^{+0.0019} $ &  $ 4.5556_{-0.0016}^{+0.0016} $ \\ 
Effective temperature, primary (K) & $5913\pm 130$ & $5606\pm 150$ \\
Effective temperature, secondary (K) & $5867\pm 130$ & $5202\pm 100$ \\
Bolometric luminosity, primary ($L_{\odot}$) & $1.49\pm0.13$       & $0.94\pm 0.10$        \\
Bolometric luminosity, secondary ($L_{\odot}$) & $1.28\pm0.11$       & $0.41\pm 0.03$        \\
{\rm [m/H]}  (dex)   & $-0.07\pm 0.15$  &  $-0.34\pm 0.20$  \\
Spectroscopic flux ratio $F_B/F_A$ (5050-5360~\AA) & $0.900\pm 0.005$ & $0.377\pm 0.015$ \\ 
\hline \hline
\multicolumn{3}{c}{{\bf Other Model Parameters}} \\ \hline 
Flux ratio in the Kepler bandpass, $F_B/F_A$ & $ 0.8475_{-0.0076}^{+0.0110} $ &  $ 0.3941_{-0.0010}^{+0.0011} $ \\ 
Primary linear limb darkening coefficient, $u_1$ & $ 0.435_{-0.040}^{+0.040} $ &  $ 0.306_{-0.051}^{+0.050} $ \\ 
Primary quadratic limb darkening coefficient, $u_2$ & $ 0.092_{-0.099}^{+0.099} $ &  $ 0.310_{-0.098}^{+0.100} $ \\ 
Secondary linear limb darkening coefficient, $u_1$ & $ 0.360_{-0.025}^{+0.026} $ &  $ 0.074_{-0.088}^{+0.087} $ \\ 
Secondary quadratic limb darkening coefficient, $u_2$ & $ 0.248_{-0.067}^{+0.064} $ &  $ 0.901_{-0.154}^{+0.155} $ \\  
Extra flux Q$_1$, $F_{X,Q_1}/F_A$ &  $ 0.0189_{-0.0034}^{+0.0035} $ &  $ 0.0630_{-0.0023}^{+0.0024} $ \\ 
Extra flux Q$_2$, $F_{X,Q_2}/F_A$ &  $ 0.0123_{-0.0034}^{+0.0035} $ &  $ 0.0706_{-0.0023}^{+0.0023} $ \\ 
Extra flux Q$_3$, $F_{X,Q_3}/F_A$ &  $ 0.0092_{-0.0034}^{+0.0035} $ &  $ 0.0662_{-0.0024}^{+0.0023} $ \\ 
Extra flux Q$_4$, $F_{X,Q_4}/F_A$ &  $ 0.0139_{-0.0034}^{+0.0035} $ &  $ 0.0407_{-0.0023}^{+0.0022} $ \\ 
Extra flux Q$_5$, $F_{X,Q_5}/F_A$ &  $ 0.0191_{-0.0034}^{+0.0035} $ &  $ 0.0620_{-0.0023}^{+0.0023} $ \\ 
Extra flux Q$_6$, $F_{X,Q_6}/F_A$ &  $ 0.0124_{-0.0034}^{+0.0035} $ &  $ 0.0682_{-0.0023}^{+0.0023} $ \\ 
Extra flux Q$_7$, $F_{X,Q_7}/F_A$ &  $ 0.0123_{-0.0034}^{+0.0035} $ &  $ 0.0668_{-0.0024}^{+0.0023} $ \\ 
Extra flux Q$_8$, $F_{X,Q_8}/F_A$ &  $ 0.0142_{-0.0034}^{+0.0035} $ &  $ 0.0387_{-0.0023}^{+0.0023} $ \\ 
Primary RV error scaling, $\sigma_A$ & $ 1.4_{-0.2}^{+0.3} $ &  $ 2.2_{-0.5}^{+0.8} $ \\ 
Secondary RV error scaling, $\sigma_B$ & $ 2.8_{-0.4}^{+0.5} $ &  $ 2.5_{-0.5}^{+0.7} $ \\ 
Photometric noise width, $\sigma_{\rm phot}$ &  $ 0.0005014_{-0.0000068}^{+0.0000068} $ &  $ 0.000848_{-0.000015}^{+0.000015} $ \\ \hline
 \end{tabular}
	}
\caption{$\vert$ Summary the stellar properties
from the output of the 
photodynamical code and TODCOR analysis, and a summary
of other model parameters for Kepler-34 and Kepler-35.
\label{tab2}}
\end{table}

\newpage

\begin{table}
\begin{centering}
{\small
\begin{tabular}{lllrrl}
\hline
Date       & UT Time & HJD & RV$_{A}$  & RV$_{B}$  & telescope \\
YYYY-MM-DD &         & (2,400,000+) & km s$^{-1}$& km s$^{-1}$ &  \\
\hline
\hline
2011-09-02&  08:35:59&  55806.8623554&$ 34.533\pm 0.057$&$-26.028\pm  0.069$&   Keck HIRES\\
2011-09-05&  11:58:33&  55810.0041793&$ 57.981\pm 0.050$&$-49.813\pm  0.056$&   Keck HIRES\\
2011-09-06&  11:47:10&  55810.9968344&$ 64.177\pm 0.044$&$-56.131\pm  0.048$&   Keck HIRES\\
2011-09-10&  07:48:37&  55814.8310956&$-25.063\pm 0.049$&$ 35.607\pm  0.052$&   Keck HIRES\\
2011-09-07&  03:33:11&  55811.6711428&$ 65.097\pm 0.165$&$-56.060\pm  0.174$&   HJST Tull\\
2011-09-08&  02:54:08&  55812.6440094&$ 50.196\pm 0.183$&$-41.578\pm  0.222$&   HJST Tull\\
2011-09-10&  03:01:17&  55814.6489276&$-21.195\pm 0.164$&$ 31.873\pm  0.207$&   HJST Tull\\
2011-09-11&  02:49:58&  55815.6410401&$-34.959\pm 0.189$&$ 44.982\pm  0.254$&   HJST Tull\\
2011-10-04&  04:34:55&  55838.7132290&$ 64.332\pm 0.129$&$-55.151\pm  0.149$&   HJST Tull\\
2011-10-06&  02:58:51&  55840.6464517&$ 43.603\pm 0.186$&$-33.956\pm  0.198$&   HJST Tull\\
2011-10-07&  04:15:40&  55841.6997478&$  4.945\pm 3.000$&$  4.945\pm  3.000$&   HJST Tull\\
2011-09-12&  06:23:36&  55816.7766219&$-39.052\pm 0.300$&$ 48.293\pm  0.275$&   HET HRS\\
2011-09-13&  06:14:15&  55817.7719168&$-37.806\pm 0.155$&$ 47.505\pm  0.184$&   HET HRS\\
2011-09-14&  05:12:48&  55818.7297401&$-35.704\pm 0.200$&$ 44.505\pm  0.255$&   HET HRS\\
2011-09-19&  05:12:48&  55823.7296067&$-17.160\pm 0.075$&$ 26.103\pm  0.086$&   HET HRS\\
2011-09-24&  04:45:32&  55828.7105309&$  4.077\pm 3.000$&$  4.077\pm  3.000$&   HET HRS\\
2011-09-25&  04:25:41&  55829.6967155&$  4.309\pm 3.000$&$  4.309\pm  3.000$&   HET HRS\\
2011-09-26&  04:38:40&  55830.7056930&$ 12.076\pm 0.090$&$ -2.822\pm  0.205$&   HET HRS\\
2011-10-08&  03:02:22&  55842.6488201&$-24.218\pm 0.253$&$ 36.070\pm  0.456$&   HJST Tull\\
2011-10-10&  04:39:10&  55844.7159536&$-37.802\pm 0.298$&$ 51.318\pm  0.287$&   HJST Tull\\
2011-10-11&  02:50:56&  55845.6407652&$-36.539\pm 0.205$&$ 48.268\pm  0.220$&   HJST Tull\\
2011-10-12&  02:47:24&  55846.6382723&$-33.893\pm 0.185$&$ 45.189\pm  0.232$&   HJST Tull\\
\hline
\end{tabular}
}
\end{centering}
\caption{$\vert$ The radial velocities for Kepler 34.  
\label{RV2459}}
\end{table}

\newpage

\begin{table}
\begin{centering}
{\small
\begin{tabular}{lllrrl}
\hline
Date       & UT Time & HJD & RV$_{A}$  & RV$_{B}$  & telescope \\
YYYY-MM-DD &         & (2,400,000+) & km s$^{-1}$& km s$^{-1}$ &  \\
\hline
\hline
2011-09-02& 09:38:57& 55806.9069580&  $   35.322\pm 0.075$ &   $    8.632\pm 0.148$ & Keck HIRES \\
2011-09-05& 12:11:10& 55810.0125860&  $   62.141\pm 0.051$ &   $  -20.583\pm 0.100$ & Keck HIRES \\
2011-09-06& 11:58:15& 55811.0044190&  $   66.440\pm 0.046$ &   $  -25.141\pm 0.086$ & Keck HIRES \\
2011-09-10& 08:00:05& 55814.8388200&  $   42.186\pm 0.055$ &   $    1.286\pm 0.115$ & Keck HIRES \\
2011-10-09& 08:40:49& 55843.8644760&  $   -8.227\pm 0.099$ &   $   57.372\pm 0.162$ & Keck HIRES \\
2011-10-16& 06:51:55& 55850.7886130&  $   57.422\pm 0.071$ &   $  -15.160\pm 0.137$ & Keck HIRES \\
2011-10-17& 07:56:14& 55851.8332030&  $   63.931\pm 0.066$ &   $  -21.988\pm 0.130$ & Keck HIRES \\
2011-10-25& 02:28:11& 55859.6190952&  $  -10.447\pm 0.093$ &   $   60.846\pm 0.202$ & HET HRS \\
2011-10-23& 19:37:37& 55858.3392335&  $    7.137\pm 0.176$ &   $   41.105\pm 0.352$ & NOT FIES \\
2011-10-25& 19:43:36& 55860.3451372&  $  -16.345\pm 0.224$ &   $   67.745\pm 0.387$ & NOT FIES \\
2011-10-26& 19:40:22& 55861.3429667&  $  -20.278\pm 0.189$ &   $   72.235\pm 0.334$ & NOT FIES \\
2011-10-29& 02:52:12& 55863.6344314&  $  -13.903\pm 0.131$ &   $   65.182\pm 0.246$ & HET HRS \\
2011-10-30& 02:03:25& 55864.6017025&  $   -7.053\pm 0.128$ &   $   57.578\pm 0.218$ & HET HRS \\
\hline
\end{tabular}
}
\end{centering}
\caption{$\vert$ The radial velocities for Kepler-35.  
\label{RV2937}}
\end{table}

\newpage

\begin{table}
\begin{tabular}{lr}
\hline
Effect   & Amplitude (ppm)\\
\hline
\hline
Beaming     &$ 214.0\pm5.7$    \\
Ellipsoidal    &$    7.1\pm 6.2$    \\
 Reflection   &$  15.8 \pm 12.4$  \\
\hline
\end{tabular}
\caption{$\vert$ The best-fit coefficients for the Doppler
beaming, the ellipsoidal effect, and the reflection effect.
\label{table_coeff}}
\end{table}

\newpage

\begin{table}
\begin{centering}
\begin{tabular}{rrrrrr}
\hline
\hline
cycle &  primary time &   error    &    cycle     &      secondary time  & error\\
      & (BJD-2,455,000)&  (minutes)&              & (BJD-2,455,000)    & (minutes) \\
\hline
0.0 &    -20.276839 &        0.100 &    0.6206712 &      -30.820781 &     0.287 \\
1.0 &      7.518920 &        0.100 &    1.6206712 &       -3.024991 &     0.834 \\
2.0 &     35.314540 &        0.105 &    2.6206712 &       24.770920 &     0.292 \\
3.0 &           ... &          ... &    3.6206712 &       52.566690 &     0.293 \\
4.0 &     90.906190 &        0.105 &    4.6206712 &       80.362200 &     0.292 \\
5.0 &    118.702150 &        0.100 &    5.6206712 &      108.158140 &     0.293 \\
6.0 &    146.497970 &        0.097 &    6.6206712 &      135.953650 &     0.295 \\
7.0 &    174.293680 &        0.097 &    7.6206712 &      163.749340 &     0.299 \\
8.0 &    202.089450 &        0.152 &    8.6206712 &      191.545570 &     0.293 \\
9.0 &           ... &          ... &    9.6206712 &      219.340980 &     0.287 \\
10.0 &     257.681310 &      0.102 &   10.6206712 &      247.136640 &     0.280 \\
11.0 &     285.476980 &      0.118 &   11.6206712 &             ... &       ... \\
12.0 &     313.272740 &      0.112 &   12.6206712 &      302.728070 &     0.304 \\
13.0 &     341.068470 &      0.100 &   13.6206712 &      330.524250 &     0.308 \\
14.0 &     368.864370 &      0.097 &   14.6206712 &      358.319840 &     0.287 \\
15.0 &     396.660230 &      0.105 &   15.6206712 &      386.115630 &     0.292 \\
16.0 &     424.456090 &      0.102 &   16.6206712 &      413.911300 &     0.290 \\
17.0 &     452.251810 &      0.100 &   17.6206712 &      441.706810 &     0.290 \\
18.0 &     480.047500 &      0.102 &   18.6206712 &      469.503040 &     0.287 \\
19.0 &     507.843500 &      0.118 &   19.6206712 &      497.298990 &     0.290 \\
20.0 &     535.639370 &      0.097 &   20.6206712 &      525.093670 &     0.287 \\
21.0 &            ... &        ... &   21.6206712 &             ... &       ... \\
22.0 &     591.230750 &      0.107 &   22.6206712 &      580.685870 &     0.287 \\
23.0 &     619.026530 &      0.100 &   23.6206712 &      608.481610 &     0.287 \\
\hline
\end{tabular}
\end{centering}
\caption{$\vert$ Times of primary and secondary eclipse for Kepler 34.  
\label{ETV34}}
\end{table}

\newpage

\begin{table}
\begin{centering}
\begin{tabular}{rrrrrr}
\hline
\hline
cycle &  primary time &   error    &    cycle     &      secondary time  & error\\
      & (BJD-2,455,000)&  (minutes)&              & (BJD-2,455,000)    & (minutes) \\
\hline
 0.0 &  -34.154064 &  0.266 &  0.5055686 &   -23.672016 &   0.312 \\
 1.0 &  -13.420444 &  0.379 &  1.5055686 &    -2.937986 &   0.256 \\
 2.0 &    7.313300 &  0.280 &  2.5055686 &    17.795530 &   0.256 \\
 3.0 &   28.046980 &  0.252 &  3.5055686 &    38.529370 &   0.256 \\
 4.0 &   48.780790 &  0.238 &  4.5055686 &    59.263080 &   0.256 \\
 5.0 &   69.514510 &  0.280 &  5.5055686 &    79.996700 &   0.270 \\
 6.0 &   90.248370 &  0.280 &  6.5055686 &          ... &     ... \\
 7.0 &  110.982030 &  0.252 &  7.5055686 &   121.464290 &   0.242 \\
 8.0 &  131.716270 &  0.294 &  8.5055686 &   142.198020 &   0.242 \\
 9.0 &  152.449660 &  0.294 &  9.5055686 &   162.931640 &   0.242 \\
10.0 &  173.183370 &  0.252 & 10.5055686 &          ... &     ... \\
11.0 &  193.917030 &  0.394 & 11.5055686 &   204.399010 &   0.396 \\
12.0 &  214.650800 &  0.394 & 12.5055686 &   225.132700 &   0.382 \\
13.0 &  235.384540 &  0.365 & 13.5055686 &   245.866550 &   0.396 \\
14.0 &  256.118370 &  0.337 & 14.5055686 &   266.600460 &   0.326 \\
15.0 &  276.852110 &  1.413 & 15.5055686 &   287.334130 &   0.242 \\
16.0 &  297.585840 &  0.294 & 16.5055686 &          ... &     ... \\
17.0 &  318.319520 &  0.252 & 17.5055686 &   328.801320 &   0.256 \\
18.0 &  339.053330 &  0.266 & 18.5055686 &   349.535060 &   0.256 \\
19.0 &  359.787070 &  0.280 & 19.5055686 &   370.268980 &   0.256 \\
20.0 &  380.520990 &  0.322 & 20.5055686 &   391.002590 &   0.242 \\
21.0 &  401.254660 &  0.280 & 21.5055686 &   411.736400 &   0.256 \\
22.0 &  421.988360 &  0.280 & 22.5055686 &          ... &     ... \\
23.0 &  442.722020 &  0.337 & 23.5055686 &   453.203900 &   0.242 \\
24.0 &  463.455920 &  0.465 & 24.5055686 &   473.937500 &   0.229 \\
25.0 &  484.189590 &  0.294 & 25.5055686 &   494.671440 &   0.256 \\
26.0 &  504.923210 &  0.252 & 26.5055686 &   515.405080 &   0.256 \\
27.0 &  525.657110 &  0.238 & 27.5055686 &   536.138860 &   0.256 \\
28.0 &  546.390840 &  0.280 & 28.5055686 &          ... &     ... \\
29.0 &         ... &    ... & 29.5055686 &   577.606230 &   0.425 \\
30.0 &  587.858380 &  0.365 & 30.5055686 &   598.339990 &   0.368 \\
31.0 &  608.592170 &  0.394 & 31.5055686 &   619.073750 &   0.396 \\
32.0 &  629.325730 &  0.379 & 32.5055686 &          ... &     ... \\
\hline
\end{tabular}
\end{centering}
\caption{$\vert$ Times of primary and secondary eclipse for Kepler 35.  
\label{ETV35}}
\end{table}

\newpage

\newpage


\begin{thebibliography}{1}



%
%
%
%
%
%
%
%
%
%
%
%

\bibitem[23]{tull1998}
Tull, R.G. High-resolution fiber-coupled spectrograph of 
the Hobby-Eberly Telescope.
{\em Proc.\ Soc.\ Photo-opt.\ Inst.\ Eng.} {\bf 3355}, 
387-398 (1998).

\bibitem[24]{endl2011}
Endl, M.,
{\em et al.}
The First Kepler Mission Planet Confirmed With The Hobby-Eberly Telescope: 
Kepler-15b, a Hot Jupiter Enriched In Heavy Elements.
{\em Astrophys.\ J.} {\bf 197} (Suppl.), 13 (2011). 

\bibitem[25]{tull1995}
{{Tull}, R.~G.,
{MacQueen}, P.~J.,
{Sneden}, C. \& {Lambert}, D.~L.}
The high-resolution cross-dispersed echelle white-pupil spectrometer 
of the McDonald Observatory 2.7-m telescope.
{\em Publ. Astron. Soc. Pacif.} {\bf 107}, 251-264 (1995).

\bibitem[26]{djupvik2010}
{{Djupvik}, A.~A. \& {Andersen}, J.}
The Nordic Optical Telescope.
in {\em Highlights of Spanish Astrophsics V}, eds.\
{J.~M.~Diego, L.~J.~Goicoechea, J.~I.~Gonz{\'a}lez-Serrano, \& 
  J.~Gorgas}, (Springer Verlag, Berlin, 2010) 211-218.

\bibitem[27]{buchhave2010}
{Buchhave}, L.~A.,
{\em et al.}
HAT-P-16b: A 4 M $_{J}$ Planet Transiting a Bright 
Star on an Eccentric Orbit.
{\em Astroph.\ J.} {\bf 720}, 1118-1125 (2010).

\bibitem[28]{vogt1994}
{Vogt}, S.~S.,
{\em et al.}
HIRES: the high-resolution echelle spectrometer on the Keck 10-m Telescope.
{\em Proc.\ Soc.\ Photo-opt.\ Inst.\ Eng.} 
{\bf 2198}, 362 (1994).

\bibitem[29]{marcy2008}
{Marcy}, G.~W.,
{\em et al.}
Exoplanet properties from Lick, Keck and AAT.
{\em Physica Scripta Volume T}, {\bf 130}, 014001 (2008).

\bibitem[30]{rucinski1992}
Recinsky, S. M.  
Spectral-line broadening functions of WUMa-type binaries. I - AW UMa.
{\em Astron.\ J.} {\bf 104}, 1968-1981 (1992).

\bibitem[31]{Nidever_2002}
{Nidever}, D.~L., {Marcy}, G.~W., {Butler}, R.~P., {Fischer}, D.~A., \& {Vogt},
  S.~S. 
Radial Velocities for 889 Late-Type Stars.
{\em Astrophys.\ J.} {\bf 141} (Suppl.), 502-522 (2002).



\bibitem[32]{soz07}
{{Sozzetti}, A.,
{Torres}, G.,
{Charbonneau}, D.,
{Latham}, D.~W.,
{Holman}, M.~J.,
{Winn}, J.~N.,
{Laird}, J.~B. \& {O'Donovan}, F.~T.}
Improving Stellar and Planetary Parameters of Transiting 
Planet Systems: The Case of TrES-2.
{\em Astrophy. J.} {\bf 664}, 1190-1198 (2007).

\bibitem[33]{zuc94}
Zucker, S. \& Mazeh, T. 
Study of spectroscopic binaries with TODCOR. 1: A new two-dimensional 
correlation algorithm to derive the radial velocities 
of the two components.
{\em Astroph. J.} {\bf 420}, 806-810 (1994).

\bibitem[34]{kur05}
Kurucz, R. L. 
ATLAS12, SYNTHE, ATLAS9, WIDTH9, et cetera.
{\em Mem.~Soc.~Astron.~Italiana, Suppl.} {\bf 8},
14-24 (2005).


\bibitem[35]{schlaufman2010}
Schlaufman, K. Evidence of Possible Spin-orbit Misalignment Along the
Line of Sight in Transiting Exoplanet Systems.
{\em Astrophys. J.}, {\bf 719}, 602-611
(2010).

\bibitem[36]{hut1981}
Hut, P. Tidal evolution in close binary systems. 
{\em Astron.\ Astrophys.} {\bf 99}, 126-140 (1981).


\bibitem[37]{loeb2003} 
Loeb, A., \& Gaudi, B.~S. 
Periodic Flux Variability of Stars due to the Reflex Doppler Effect 
Induced by Planetary Companions.
{\em Astrophys. J.} {\bf 588}, L117-L120 (2003).

\bibitem[38]{zucker2007} 
Zucker, S., Mazeh, T., \& Alexander, T.
Beaming Binaries: A New Observational Category of 
Photometric Binary Stars.
{\em Astrophys. J.} {\bf 670}, 1326-1330 (2007). 

\bibitem[39]{rybicki1979}
Rybicki, G. B., \& Lightman, A. P.
{\em Radiative Processes in Astrophysics} 
(Wiley-Interscience, New York 1979).

\bibitem[40]{mazeh2008} 
Mazeh, T.
Observational Evidence for Tidal Interaction in Close Binary Systems. 
{\em EAS 
Publications Series} {\bf 29}, 1-65 (2008). 

\bibitem[41]{for2010} 
{For}, B.-Q.,
{\em et al.}
Modeling the System Parameters of 2M 1533+3759: 
A New Longer Period Low-Mass Eclipsing sdB+dM Binary.
{\em Astrophys. J.} {\bf 708}, 253-267 (2010).

\bibitem[42]{mazeh2010} 
Mazeh, T., \& Faigler, S.
Detection of the ellipsoidal and the relativistic 
beaming effects in the CoRoT-3 lightcurve.
{\em Astron.\ Astrophys.} {\bf 521}, L59-L63 (2010).

\bibitem[43]{faigler2011} 
Faigler, S., \& Mazeh, T.
Photometric detection of non-transiting short-period 
low-mass companions through the beaming, ellipsoidal and reflection 
effects in Kepler and CoRoT light curves.
{\em Mot.\ Not.\ R. Astron.\ Soc.} {\bf 415}, 3921-3928 (2011).


\bibitem[44]{demory2011} 
{Demory}, B.-O.,
{\em et al.}
The High Albedo of the Hot Jupiter Kepler-7 b.
{\em Astrophys. J.} {\bf 735}, L12-L18 (2011).



\bibitem[45]{faigler2011b} 
Faigler, S., Mazeh, T., 
Quinn, S.~N., Latham, D.~W., \& Tal-Or, L.
Seven new binaries discovered in the Kepler light curves through the 
BEER method confirmed by radial-velocity observations.
{\em Astrophys. J.} submitted,
arXiv:1110.2133 (2011).


\bibitem[46]{steffen2011}
{Steffen}, J.~H.,
{\em et al.}
The architecture of the hierarchical triple star KOI 928 from 
eclipse timing variations seen in Kepler photometry.
{\em Mot.\ Not.\ R. Astron.\ Soc.} {\bf 417}, L31-L35 (2011).




\bibitem[47]{soderhjelm1984}
Soderhjelm, S. 
Third-order and tidal effects in the stellar three-body problem.
{\em Astron.\ Astroph.} {\bf 141}, 232-240 (1084).

\bibitem[48]{mardling2002}
Mardling, R. \& Lin, D. N. C.
Calculating the Tidal, Spin, and Dynamical Evolution of 
Extrasolar Planetary Systems.
{\em Astrophys. J.} {\bf 573}, 829-844 (2002).

\bibitem[49]{soffel1989}
Soffel, M. H. {\em Relativity in Astrometry,
Celestial Mechanics ande Geodesy XIV},
(Springer-Verlag, Berlin, 1989).

\bibitem[50]{press2007}
Press, W. H., Teukolsky, S. A., Vetterling, W. T., \&
Flannery, B. P.
{\em Numerical Recipes in C++}, (Cambridge University Press,
Cambridge, 2007).

\bibitem[51]{limb}
The tabulated limb-darkening coefficients are available at \\
http://astro4.ast.villanova.edu/aprsa/?q=node/8.

\bibitem[52]{braak2006}
Braak, C. J. F.
A Markov Chain Monte Carlo Version of the Genetic Algorithm
Differential Evolution: Easy Bayesian Computing for Real Parameter
Space.  {\em Stat.\ Comput.} {\bf 16}, 239 (2006).



\bibitem[53]{allard}
See, for example, computations by F. Allard
at \\ http://phoenix.ens-lyon.fr/Grids/NextGen/COLORS/colmag.NextGen.server.2MASS.






\bibitem[54]{press1992}
Press, W. H., Teukolsky, S. A., Vetterling, W. T., \&
Flannery, B. P.
{\em Numerical Recipes in Fortran}, (Cambridge University Press,
Cambridge, 1992).

\bibitem[55]{mercury}
The software is available at
http://www.arm.ac.uk/$\sim$jec/home.html.

\bibitem[56]{kokubo1998}
Kokubo, E., 
Yoshingaga, K., \&
Makino, J. On a time-symmetric Hermite integrator for planetary N-body simulation.
{\em Mot.\ Not.\ R. Astron.\ Soc.} {\bf 297}, 1067-1072 (1998).

\bibitem[57]{swarm}
The software is available at
http://www.astro.ufl.edu/$\sim$eford/code/swarm/.





\bibitem[58]{Showman02}
{Showman}, A.~P. \& {Guillot}, T. 
Atmospheric circulation and tides of ``51 Pegasus b-like'' planets.
{\em Astronom.\ Astrophy.} {\bf 385}, 166-180 (2002).


\bibitem[59]{Showman10}
{Showman}, A.~P., {Cho}, J.~Y.~-K., \& 
{Menou}, K. 
Atmospheric Circulation of Exoplanets.
in {\em Exoplanets}, ed.~S.~Seager, (University of Arizona Press,
Tucson, 2010).





\bibitem[60]{xue2008}
{Xue}, X.~X.,
{\em et al.}
The Milky Way's Circular Velocity Curve to 60 kpc and
an Estimate of the Dark Matter Halo Mass from the Kinematics of
$\sim 2400$ SDSS Blue Horizontal-Branch Stars.
{\em Astrophys.\ J.} {\bf 684}, 1143-1158 (2008).


\bibitem[61]{cumming2008}
{Cumming}, A.,
{Butler}, R.~P.,
{Marcy}, G.~W.,
{Vogt}, S.~S.,
{Wright}, J.~T., \&
{Fischer}, D.~A.
The Keck Planet Search: Detectability and the Minimum Mass and Orbital Period 
Distribution of Extrasolar Planets.
{\em Pub.\ Astron.\ Soc.\ Pacif.} {\bf 120}, 531-554 (2008).

\bibitem[62]{binney}
Binney, J. \& Tremaine, S.
{\em Galactic Dynamics, Second Edition},
(Cambridge University Press, Cambridge, 2008).

\end{thebibliography}
\end{document}